\pdfoutput=1
\documentclass[a4paper,10pt,twocolumn,amsmath,amssymb,nofootinbib,%
superscriptaddress,groupedaddress,floats,floatfix,aps,prd]{revtex4-1}
\usepackage[british]{babel}
\usepackage[latin1]{inputenc}
\usepackage[T1]{fontenc}
\usepackage{graphicx,color,soul}
\usepackage[pdftex,pdfusetitle]{hyperref}
\hypersetup{colorlinks=true,linkcolor=red,citecolor=blue,filecolor=black,urlcolor=black,
pdfauthor={Edgardo Franzin, Mariano Cadoni and Matteo Tuveri}}
\usepackage[capitalize]{cleveref}

\def\be#1\ee{\begin{align}#1\end{align}}

\newcommand{\ie}{i.e.}
\newcommand{\eg}{e.g.}
\newcommand{\p}{\partial}
\newcommand{\0}{\nonumber}

\renewcommand{\le}{\leqslant}
\renewcommand{\leq}{\leqslant}
\renewcommand\O{\mathcal{O}}

\DeclareMathOperator\Li{Li}

\begin{document}

\title{Sine-Gordon solitonic scalar stars and black holes}

\date{6 June 2018}

\author{Edgardo Franzin}\email[Corresponding author:~]{edgardo.franzin@ca.infn.it}
\author{Mariano Cadoni}
\author{Matteo Tuveri}
\affiliation{Dipartimento di Fisica, Universit\`a di Cagliari
\& INFN, Sezione di Cagliari\\
Cittadella Universitaria, 09042 Monserrato, Italy}

\begin{abstract}
We study exact, analytic, static, spherically symmetric, four-dimensional solutions of minimally coupled Einstein-scalar gravity, sourced by a scalar field whose profile has the form of the sine-Gordon soliton.
We present a horizonless, everywhere regular and positive-mass solution ---~a solitonic star~--- and a black hole.
The scalar potential behaves as a constant near the origin and vanishes at infinity.
In particular, the solitonic scalar star interpolates between an anti-de Sitter and an asympototically flat spacetime.
The black-hole spacetime is unstable against linear perturbations, while due to numerical issues, we were not able to determine with confidence whether or not the star-like background solution is stable.
\end{abstract}

\maketitle%

\section{Introduction}

To date, it is well-known that visible baryonic matter accounts for only a small part of the total mass of the universe.
The most reliable and conservative approach to dark matter is the $\Lambda$CDM model~\cite{Ade:2015xua}, but several alternatives have been introduced to take into account some problems present in the model --- from modifications of general relativity~\cite{Milgrom:1983ca}, to particle dark matter~\cite{Bertone:2010} and emergent gravity approaches~\cite{Verlinde:2016toy,Cadoni:2017evg,Cadoni:2018dnd}.
However, as dark matter is most likely non-baryonic, it is interesting to consider asymptotically flat self-gravitating objects made up of massive fundamental (pseudo) scalar fields and to study their astrophysical consequences~\cite{Marsh:2015wka}.

Boson stars~\cite{Schunck:2003kk,Liebling:2012fv} are the most famous example: they are non-topological solitonic configurations of massive \emph{complex} scalar fields non-linearly coupled to themselves through a self-interacting scalar potential and to gravity.
Stable and compact configurations have also been proposed as alternatives to astrophysical and primordial black holes~\cite{Mielke:2000mh,Guzman:2009zz}.
In fact, gravitational collapse could stop before the object reaches its Schwarzschild radius to produce a horizonless object that mimics some observational features of black holes~\cite{Macedo:2013jja,Barausse:2014tra,Vincent:2015xta}, but that may still be distinguished from signatures in the gravitational-wave waveform~\cite{Macedo:2013qea,Cardoso:2016rao,Cardoso:2016oxy,Cardoso:2017cfl,Sennett:2017etc}.

In boson stars, the constituent complex scalar fields are globally invariant under U(1) symmetry and, as a consequence, there exists a conserved Noether current.
For \emph{real} massive scalar fields there is no such a current and the situation is very different: there are no static solutions.
However there exist oscillatons~\cite{Seidel:1991zh}, for which both the metric and the scalar field are periodically oscillating in time.

The key observation is that boson stars and oscillatons are found by fixing the scalar potential.
Then, the metric functions and the scalar profile are determined by solving the Einstein-Klein-Gordon equations.
Here, on the contrary, we fix the scalar profile, we determine the scalar potential dynamically and we show that static regular self-gravitating solutions made up of real scalar fields are allowed.

Black-hole solutions sourced by scalar fields in asymptotically flat spacetimes are generically forbidden by no-hair theorems which relate the existence of hairy black holes to the non-convexity of the potential~\cite{Bekenstein:1971hc,Bekenstein:1995un,Sudarsky:1995zg} and to the violation of the positive energy theorem~\cite{Torii:2001pg,Hertog:2006rr} with some notable exceptions~\cite{Herdeiro:2014goa,Herdeiro:2015waa}.
In some cases, the zero-event-horizon limit describes an everywhere regular, particle-like object known as scalaron~\cite{Nucamendi:1995ex}.

In this work we study exact, analytic, static, spherically symmetric, four-dimensional solutions of minimally coupled Einstein-scalar gravity --- for some examples, see \eg\ Ref.~\cite{Bechmann:1995sa}.
We derive both a horizonless, everywhere regular, positive-mass solution and a black hole.
These solutions are sourced by a scalar field whose profile is identical to that of the sine-Gordon soliton~\cite{Rubinstein:1970vb}.
These solitons have a wide range of applications in several areas of non-linear physics, \eg\ non-linear molecular and DNA dynamics, the Josephson effect, ferromagnetic waves, non-linear optics, superconductivity and many others~\cite{Gaeta,Barone,Cuevas-Maraver}.
In two-dimensional gravity, there exists a relationship between the sine-Gordon dynamics and the black-hole metric degrees of freedom~\cite{Gegenberg:1997ns,Cadoni:1998ej}, while a sine-Gordon star is known in Brans-Dicke gravity~\cite{Su:2012}.
Thus, it is remarkable that a sine-Gordon soliton may also act as a gravitational scalar source in general relativity.

The energy density of the horizonless solution is negative close to the origin but it is balanced by a positive energy density in the asymptotic region to produce a positive total gravitational mass.
Plus, this self-gravitating configuration sourced by a sine-Gordon scalar profile has compactness of $\O(0.1)$.
For these reasons, we call it a sine-Gordon solitonic scalar star.

To derive these solutions we utilise a slightly different version of the solution-generating method proposed in Ref.~\cite{Cadoni:2011nq} which has been successfully used to obtain a large number of exact, static, asymptotically flat or anti-de Sitter (AdS) black-hole and black-brane solutions~\cite{Cadoni:2011yj,Cadoni:2012uf,Cadoni:2012ea,Cadoni:2015gfa,Cadoni:2015gce}.
We do not give details about this new version of the method here, but the essential result is that ---~under certain assumptions on the reality of the scalar field and the asymptotic behaviour of the spacetime~--- the solution is completely parametrised by a single function.
An equivalent method has been presented in Refs.~\cite{AzregAinou:2009dj,Solovyev:2012zz}.

Throughout this work we adopt $c=16\pi G=1$ units.

\section{Solitonic solutions}

We consider four-dimensional Einstein gravity minimally coupled to a self-interacting real scalar field $\phi$,
\be\label{action}
S = \int d^{4}x\,\sqrt{-g}\left(\mathcal{R} - \frac{1}{2}\,\p_\mu\phi\,\p^\mu\phi - V(\phi)\right),
\ee%
and we look for asymptotically flat, static, spherically symmetric solutions $ds^2 = - U(r)\,dt^2 + U(r)^{-1}dr^2 + R^2(r)\,d\Omega_2^2$ sourced by a scalar which inherits the spacetime symmetries~\cite{Smolic:2015txa,Smolic:2016dmh} and whose stress-energy tensor is
\be\label{Tmunu}
T_{\mu\nu} = \p_\mu\phi\,\p_\nu\phi - g_{\mu\nu}\left(\frac{1}{2}\,\p_\mu\phi\,\p^\mu\phi + V(\phi)\right).
\ee%

Introducing an auxiliary dimensionless coordinate \mbox{$x\equiv r_0/r$}, with $r_0$ arbitrary length scale --- which we will see proportional to the gravitational mass of the solution and inverse proportional to the square root of the amplitude of the scalar potential --- the solution of the field equations can be entirely parametrised by a single function $P(x)$ and can be recast in the form,
\be%
& R(x)=\frac{r_0 P}{x}\,,\quad \phi(x) = 2\int dx\,\sqrt{-\frac{1}{P}\frac{d^2 P}{dx^2}}\,,\quad\label{sol1}\\
& U(x) = \frac{r_0^2 P^2}{x^2} \left(c_2 + \frac{2}{r_0^2}\int\frac{x\,dx}{P^4} + \frac{c_1}{r_0^3}\int\frac{x^2\,dx}{P^4}\right),\label{sol2}\\
& V[\phi(x)] =\frac{x^2}{2 r_0^2 P^2}\left[2 - x^2\frac{d}{dx}\left(x^2 \frac{d}{dx}\frac{U P^2}{x^2}\right)\right],\label{sol3}
\ee%
where $c_1$ and $c_2$ are integration constants, whose value can be determined by the boundary conditions of the spacetime.

The $r$-asymptotic region corresponds to $x=0$, while the $r$-origin corresponds either to $x=\infty$ when $P(x)$ has no zeros at finite values, or to $x=x_0$ when $P(x_0)=0$.
Because of its relation with the radius $R$ of the $2$-sphere, $P(x)$ must be a positive, analytic and monotonically decreasing function.
Moreover, the condition of asymptotic flatness requires $P(0)=1$ and reality of the scalar field implies $d^2P/dx^2\le 0$.
When $P(x)$ has a zero at a finite value $x_0$, $U(x_0)$ becomes singular and in view of its integral form \eqref{sol2}, quite generically the spacetime will develop a curvature singularity.
The only way to avoid such a curvature singularity, but still have non-trivial solutions, is to impose an asymptotically constant scalar field profile and an exponential decreasing of $d^2P/dx^2$.
In fact, from the field equations it turns out that the scalar curvature is given by
\be%
\mathcal{R}= 2V - \frac{x^4 U}{r_0^2 P}\frac{d^2P}{dx^2}\,,\0
\ee%
hence, the exponential behaviour of $d^2P/dx^2$ is needed to kill the power-law divergences in $\mathcal{R}$.
The simplest choice for a function $P$ satisfying all the conditions above is
\be\label{P}
P(x)=2-e^{-x}\,.
\ee%

For the rest of the work we switch back to the radial coordinate $r$.
From \cref{sol1}, the metric function $R$ is
\be\label{Rsol}
R(r) = r\left(2-e^{-r_0/r}\right),
\ee%
and surprisingly enough, the scalar field profile turns out to be identical to that of the solitons (kinks) of the sine-Gordon theory~\cite{Rubinstein:1970vb},
\be\label{scalar0}
\phi(r) = \pi - 4\arcsin\frac{e^{-r_0/2r}}{\sqrt{2}}\,.
\ee%
The scalar field stays always finite, goes to zero asymptotically as $\phi\sim r_0/r$, whereas it behaves exponentially near the origin, \ie\ $(\phi-\pi)\sim e^{-r_0/2r}$ as $r\to0$.

Fixing the value of $c_2$ to have an asymptotically flat solution, \ie\ $U(r)\to1$ as $r\to\infty$, the metric function $U$ can be written as the sum of a regular and a divergent term in the origin, 
\be\label{generalU}
U(r) = \frac{r^2 P^2}{48r_0^2}\left( u_\text{reg}(r) + \frac{c_1}{r_0}\,u_\text{BH}(r) \right),
\ee%
with
\begin{widetext}
\be%
u_\text{reg}(r) &= \frac{a^2}{2} + \left(\frac{r_0}{r}+3\right)\left(\frac{3r_0}{r}+2\right)-\frac{16r_0}{r P^3}-\frac{4 (3r_0+r)}{r P^2}-\frac{2 (6r_0+5r)}{rP}+\left(\frac{6r_0}{r}+11\right) \log\frac{P}{2}-6 \Li_2\left(1-\frac{P}{2}\right),\\
u_\text{BH}(r) &= b^2 + \frac{r_0 (r_0+4r) (2r_0+3r)}{2r^3} -\frac{8 r_0^2}{r^2 P^3}-\frac{2r_0 (3r_0+2r)}{r^2 P^2}-\frac{2r_0 (3r_0+5r)+2r^2}{r^2 P} + 6\log{P}\0\\
&\phantom{=}+ \frac{r_0}{r} \left(\frac{3r_0}{r}+11\right) \log\frac{P}{2} - \left(\frac{6r_0}{r}+11\right) \Li_2\left(1-\frac{P}{2}\right)-6 \Li_3\left(1-\frac{P}{2}\right),
\ee%
where $a$ and $b$ are numerical constants, $a^2 = 16 + 22\log 2 + \pi^2 - 6\log^2 2$ and $b^2 = 2 + 21\zeta(3)/4 - \log^2{2}(11-2 \log2)/2 + \pi^2 (11-6 \log2)/12$.
\end{widetext}

Depending on the value and the sign of $c_1/r_0$, the metric function in \cref{generalU} describes either a black hole (discussed in \cref{bhbranch}), a naked singularity, or a regular star-like solution (discussed in \cref{starbranch}).

\begin{figure}[h]
\centering%
\includegraphics[width=0.46\textwidth]{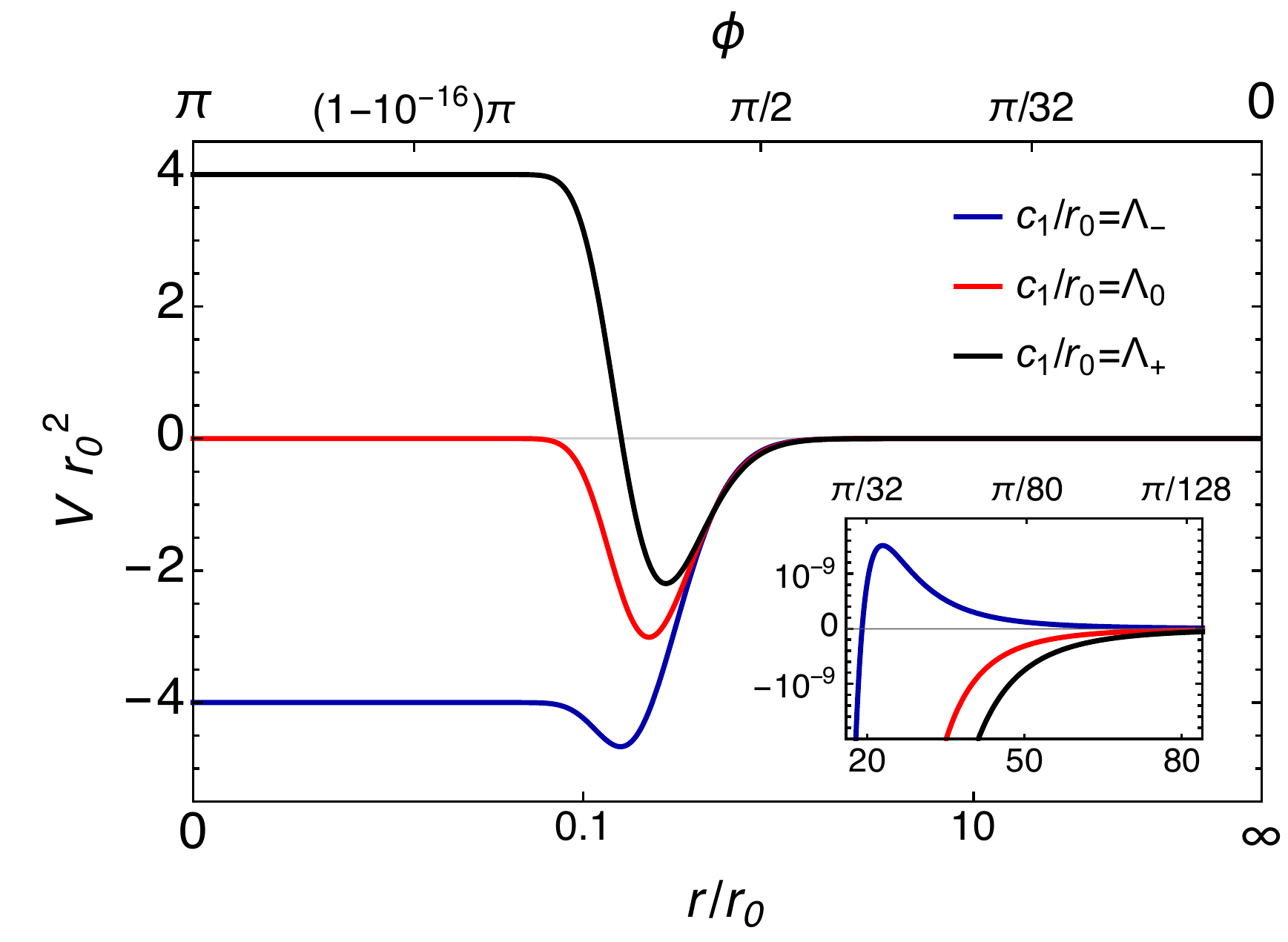}\quad%
\includegraphics[width=0.46\textwidth]{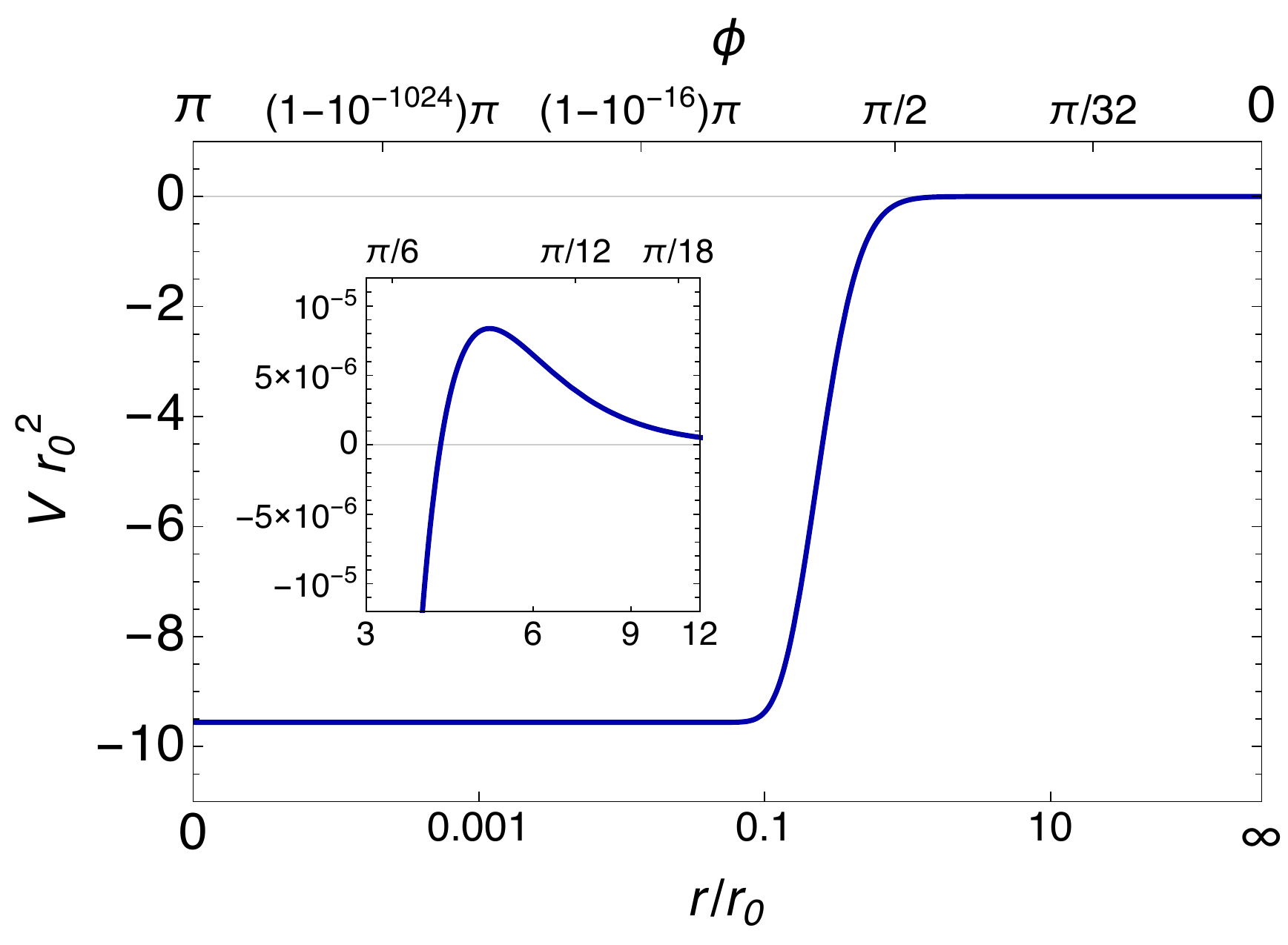}
\caption{Plots of the scalar potential as a function of $r$ and~$\phi$.
Left panel: The potential in the black-hole branch.
$\Lambda_\pm$ are such that the potential goes to $\pm4/r_0^2$ in the origin, \ie\ $\Lambda_\pm \equiv (a^2 \pm 16)/(2 - 2b^2 - 12\log2)$.
Inset: Zoom on the asymptotic region.
Right panel: The potential for the star-like solution, $c_1=0$. Inset: Zoom on the maximum.}
\label{fig:potential}
\end{figure}

The expression for the potential~\eqref{sol3} can be computed analytically but is cumbersome.
We give in \cref{fig:potential} representative plots as functions of $r$ and $\phi$, both in the black-hole and the star-like branch.
The scalar potential goes to zero asymptotically ($r\to\infty$, \ie\ $\phi\to0$) as
\be%
V(\phi\approx0) = \frac{c_1+r_0}{120 r_0^3}\,\phi^5 + \O(\phi^6)\,,\0
\ee%
while near the origin ($r\to0$ \ie\ $\phi\to \pi$), it approaches a constant
\be%
V(\phi\approx\pi) = -\frac{a^2}{4r_0^2} - \frac{c_1}{r_0} \frac{1-b^2-6\log2}{2r_0^2} + \O(\phi-\pi)\,.\0
\ee%

\onecolumngrid

\twocolumngrid

\subsection{Black-hole branch\label{bhbranch}}

For non-zero values of the integration constant $c_1$, the metric function $U$ in \cref{generalU} has a curvature singularity in $r=0$, in fact, while the curvature and Riemann scalars are finite at $r=0$, the Kretschmann scalar diverges.
It describes either a black hole ($c_1/r_0<0$) or a naked singularity ($c_1/r_0>0$).
For the rest of the work we focus on the black-hole case.

The gravitational mass $M$ of the solution can be easily inferred from the $1/r$ term in the asymptotic expansion of the metric function $U(r)$; it is positive and given by \mbox{$M=8\pi (2r_0-c_1)/3$}.
The black hole event horizon $r_H$ is defined implicitly by $U(r_H)=0$ and is always within the corresponding Schwarzschild radius.
We notice that the $c_1\to0$ limit is singular: in fact, the black hole horizon goes to zero while the black hole mass tends to the finite value $16\pi r_0/3$.

The scalar potential for the black-hole branch is plotted in the left panel of \cref{fig:potential} for representative values of $c_1/r_0$.
It always possesses a flat region near $r=0$ followed by a minimum. Then the potential vanishes asymptotically; in particular, for $c_1/r_0<-1$, it goes to zero from below.
Notice that for $c_1/r_0 = \Lambda_0\equiv -a^2/(2 b^2-2+12 \log2)$ the scalar potential is zero at the origin, while for $c_1/r_0$ greater (less) than $\Lambda_0$, the value of the constant becomes negative (positive).

We stress that the scalar potential depends on the value of $c_1/r_0$ and, as a consequence, the formulation of a consistent black-hole thermodynamics is very difficult.
In principle, one could get rid of this unpleasant feature using an appropriate rescaling of the parameters appearing in the potential, along the lines described in Ref.~\cite{Cadoni:2015gfa} for black holes sourced by massless scalars.
In the case under consideration, this is a rather involved issue because of the complicate form of the potential both as a function of the coordinate $r$ and of the scalar field $\phi$.
On the other hand, the main focus of this work is not on the black-hole branch but rather on the star-like branch and its stability.
We use therefore the black-hole case as a proxy to discuss the stability of the star-like branch in the $c_1/r_0\to0$ limit.

\subsection{Star-like branch\label{starbranch}}

When $c_1=0$, the metric function $U$ describes a horizonless and perfectly regular solution with no curvature singularities,
\be\label{Usol}
U(r) = \frac{r^2 P^2}{48r_0^2}\,u_\text{reg}(r)\,.
\ee%
We stress that the star-like branch cannot be considered as the $c_1\to0$ limit of the black-hole branch as such a limit is singular.
Near the origin, after the coordinate rescaling $r\to r/2$, the metric functions behave as $R(r)=r$ and $U(r)=r^2/L^2+1$, \ie\ it describes an AdS spacetime with AdS length $L^2 = 6 r_0^2/a^2$.

In this case, the gravitational mass of the solution is again positive and its value is $M=16\pi r_0/3$.
As the scalar field is spread all over the radial direction, this solution does not have a hard surface.
Yet we could define an effective radius $r_\text{eff}$ within which 99\% of the mass is contained. It turns out to be, roughly, $r_\text{eff}/r_0\approx98$, almost three times larger than its Schwarz\-schild radius.
This also means that the compactness of this solution is about 0.17, a value compatible with other boson and fluid stars but not black holes ---~see \eg\ Fig.~4 of Ref.~\cite{AmaroSeoane:2010qx}.

This solution represents an extremely non-trivial gravitational configuration, which we call a sine-Gordon solitonic scalar star.
The solution itself has a solitonic nature because it has a positive mass, it is completely free of spacetime singularities and it interpolates between two maximally symmetric spacetimes --- an asymptotically flat spacetime at $r=\infty$ and an AdS spacetime at $r=0$.

The scalar potential is plotted in the right panel of \cref{fig:potential} both as a function of $r$ and $\phi$.
Near the origin, it approaches a negative constant $V=-3/2L^2=-a^2/4r_0^2$ consistently with its AdS behaviour.
It is interesting to notice that the potential is positive for large values of $r$ (see the inset in \cref{fig:potential}), reaches a maximum at around $r/r_0\approx5.01$ then crosses the axis for $r/r_0\approx4.08$ and goes down to negative values to approach exponentially the constant negative AdS value.

Despite the fact that in general a scalar field does not obey an equation of state~\cite{Madsen:1985}, the stress-energy tensor of the scalar field \eqref{Tmunu} can also be interpreted as produced by a non-perfect, anisotropic fluid with both radial and perpendicular pressure,
\be%
-T^0_0 &= \rho = \frac{1}{2}\,U\phi'^2+V = \mathcal{T}+V\,,\\
T^1_1 &= p_\text{rad} = \frac{1}{2}\,U\phi'^2-V\,,\quad
T^2_2 = p_\text{tan} = -\rho\,.
\ee%

\begin{figure}[!h]
\centering
\includegraphics[width=0.22\textwidth]{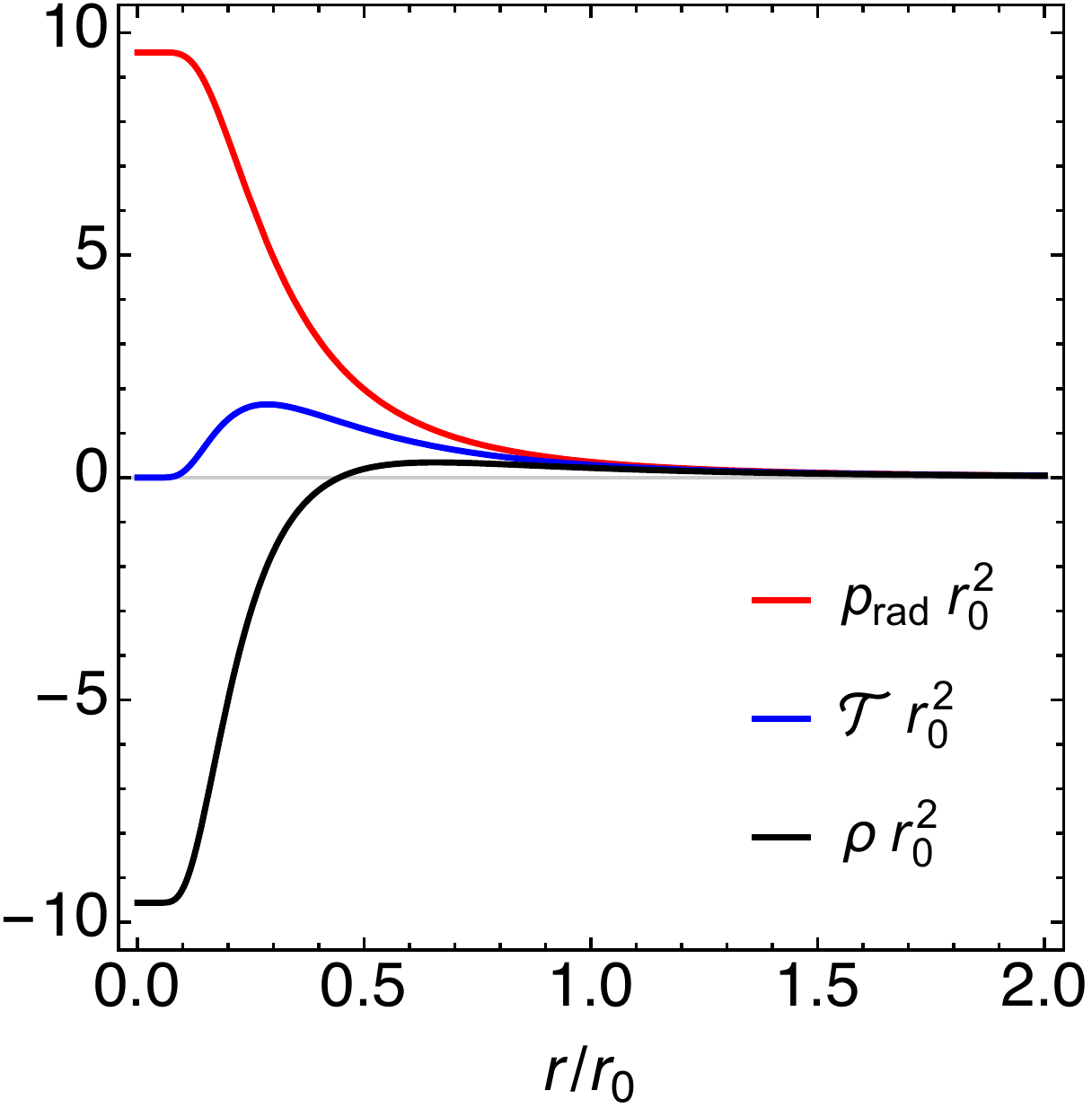}\quad
\includegraphics[width=0.22\textwidth]{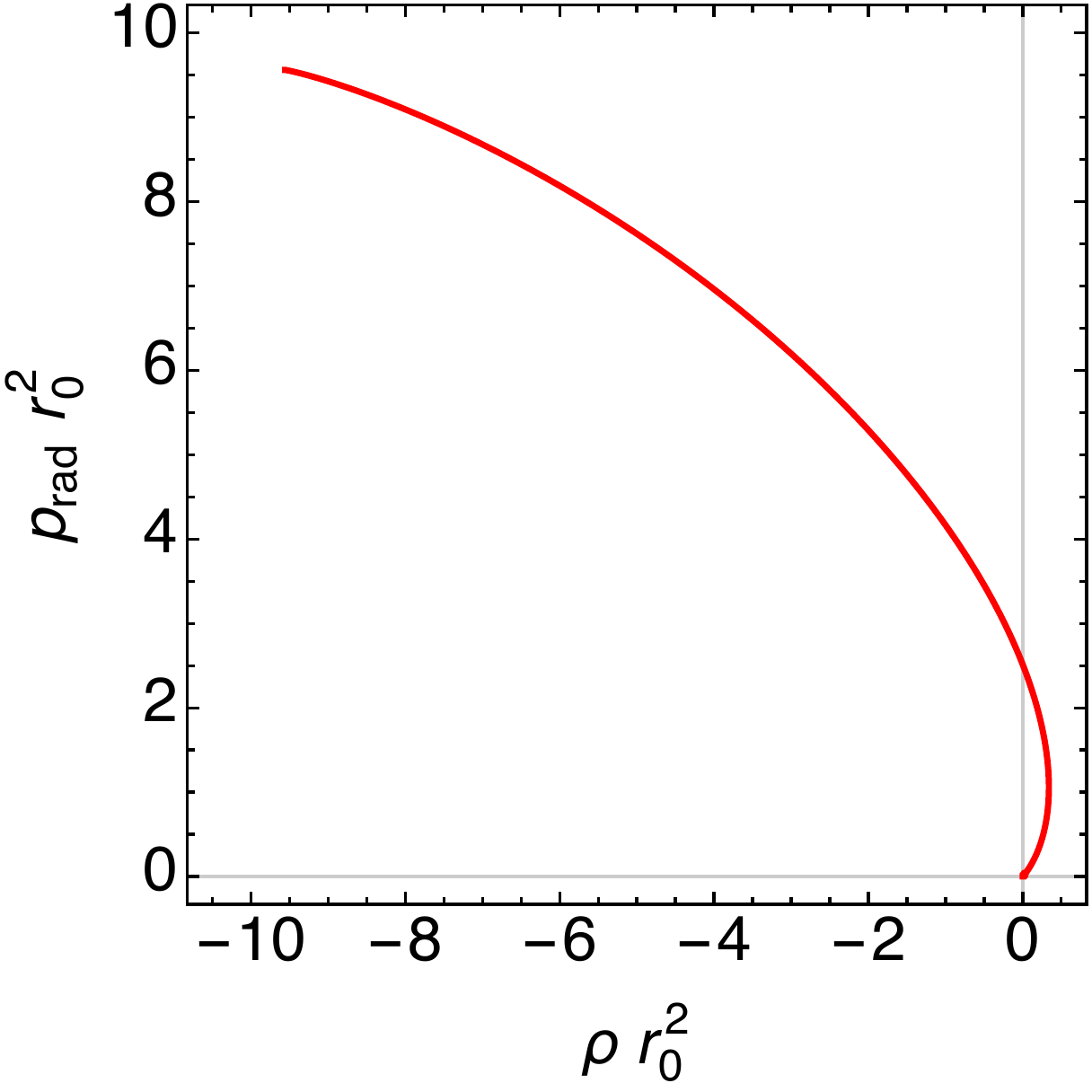}
\caption{Left panel: The energy density $\rho$, its kinetic contribution $\mathcal{T}$ and the radial pressure $p_\text{rad}$ as functions of the radial coordinate~$r$. Right panel: Equation of state.}
\label{fig:energy}
\end{figure}

In \cref{fig:energy} on the left we plot the energy density~$\rho$, its kinetic contribution~$\mathcal{T}$, and the radial pressure~$p_\text{rad}$ as functions of $r$, while on the right we plot the position-dependent equation of state $p_\text{rad}=p_\text{rad}(\rho)$.

Although the energy density is negative for small values of $r$, the gravitational mass is positive.
The existence of this positive mass solution results from the peculiar highly non-linear interaction of the scalar field producing a negative energy density in the inner region balanced by the positive energy density in the asymptotic region.
In order to see if this balance may produce a stable configuration, we have to investigate the stability of our solution.

\section{Stability analysis}

To discuss the stability of our solutions we consider $s$-wave radial perturbations (they are generically expected to be the least stable) about the background, \ie\ $U(r)+\delta U(t,r)$, $R(r)+\delta R(t,r)$ and $\phi(r)+\delta \phi(t,r)$.

By expanding the field equations up to linear order in the perturbation fields and by making use of the background equations, the perturbation equations reduce to two constraints and a dynamic equation for $\delta\phi$~\cite{Bronnikov:2011if}.

Furthermore, assuming harmonic time dependence for the scalar perturbation
\be%
\delta\phi(t,r) \equiv e^{-i\omega t} R(r) \psi(r)\,,\0
\ee%
the master equation for radial perturbations reads
\be\label{eqpert}
\frac{d^2\psi}{dr_*^2} + \left(\omega^2 - V_\text{eff}\right)\psi = 0\,,
\ee%
where $r_*$ is a ``tortoise'' coordinate%
\footnote{$r_*$ is an actual tortoise coordinate in the black-hole branch where $r\to r_H$ is mapped into $r_*\to-\infty$ and $r\to\infty$ into $r_*\to\infty$. In the star-like branch, $r\to0$ corresponds to a finite value $r_*\to r_*^0$.}
$dr_*/dr=1/U(r)$ and
\be\label{effectiveV}
\frac{V_\text{eff}}{U} = \frac{1 - U R'^2}{R^2}+\frac{\left(V R^2-2\right) \phi'^2}{4 R'^2}+\frac{V_\phi R \phi'}{R'}-\frac{V}{2}+V_{\phi\phi}\,,
\ee%
where $V_\phi=d V/d\phi$ and $V_{\phi\phi}=d^2 V/d\phi^2$.

\begin{figure*}[hbt]
\centering%
\includegraphics[width=0.485\textwidth]{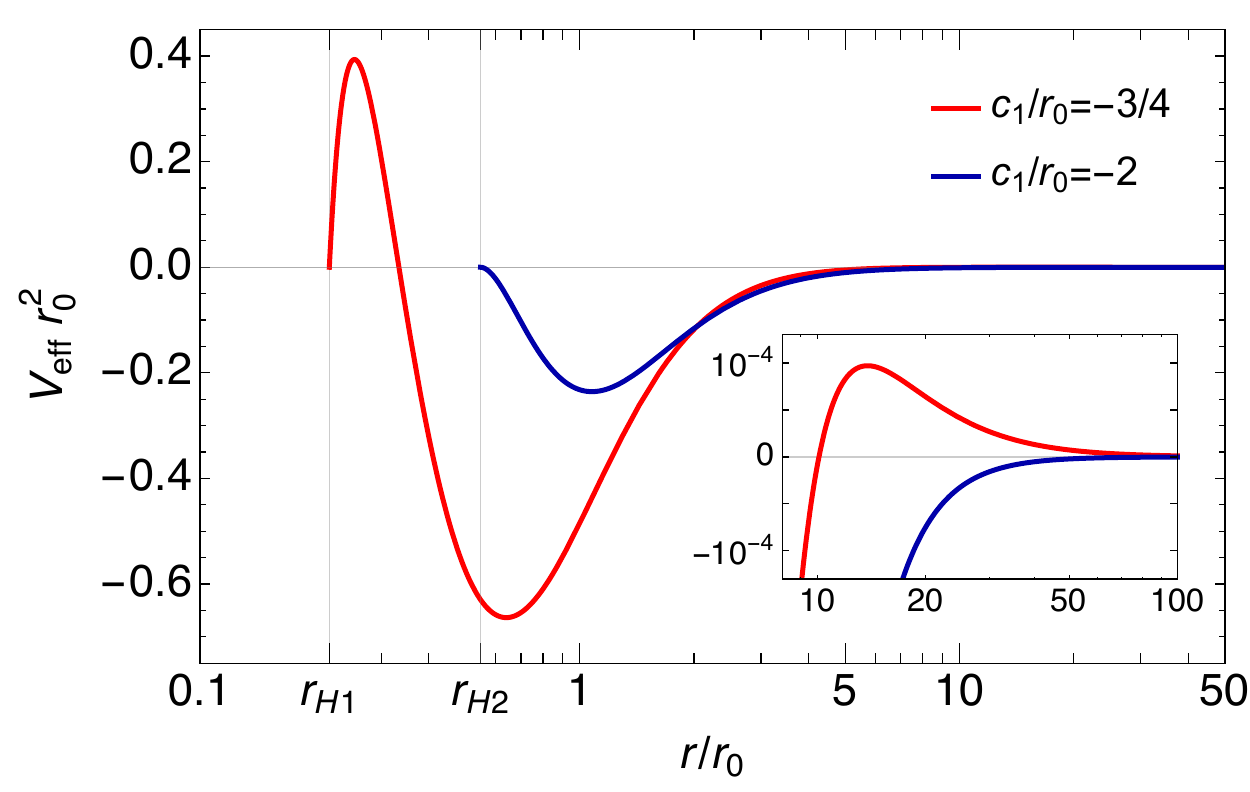}\quad%
\includegraphics[width=0.46\textwidth]{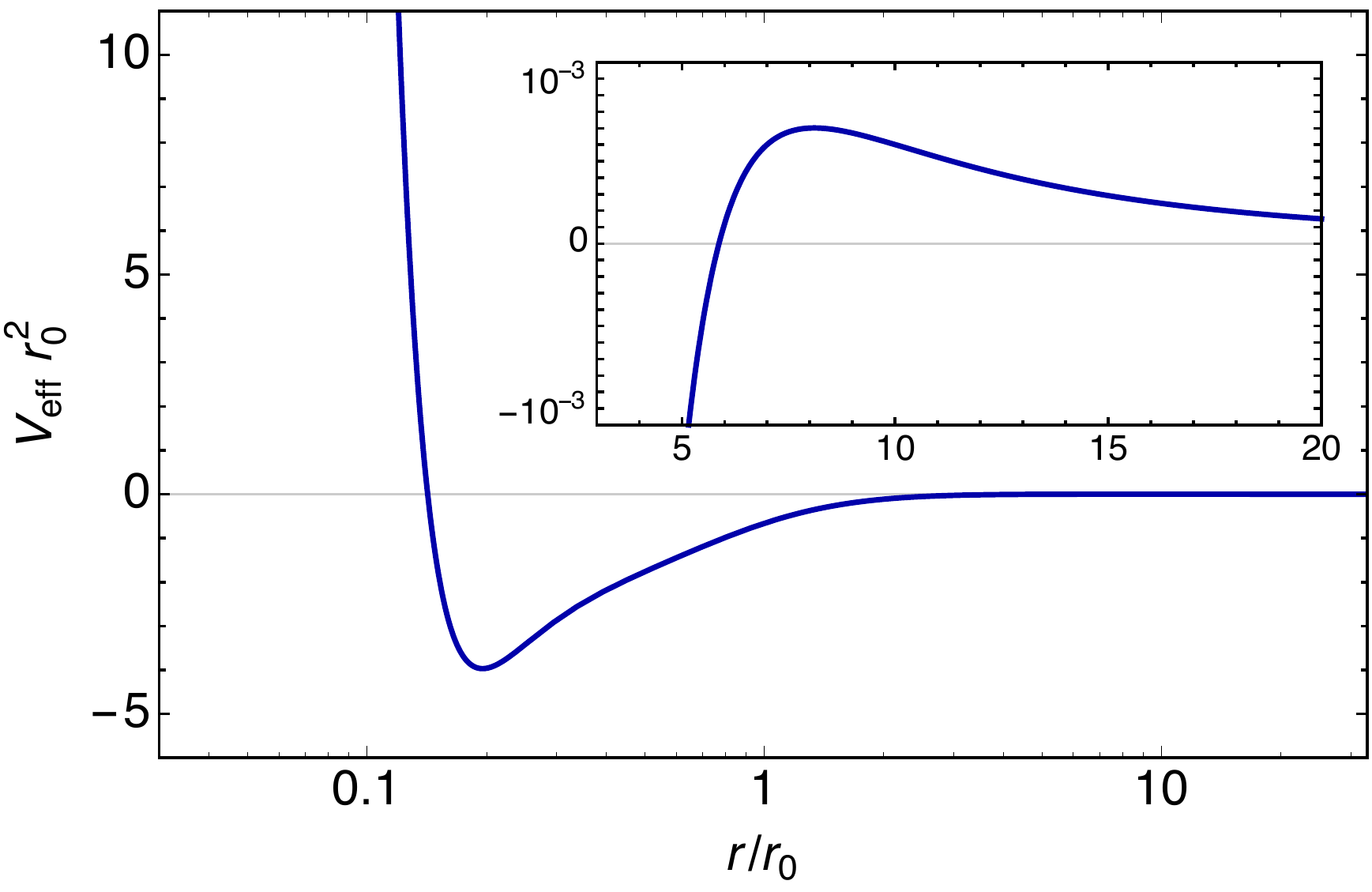}
\caption{Plots of the effective potential as a function of $r$.
Left panel: The effective potential in the black-hole branch for two representative values of $c_1/r_0$.
$r_{H1}$ and $r_{H2}$ are the two corresponding event horizons, $r_{H1}/r_0\approx0.219$ and $r_{H2}/r_0\approx0.547$.
Inset: Zoom on the secondary potential barrier.
Right panel: The effective potential for the star-like solution. Zeros for $r/r_0\approx0.14$ and $r/r_0\approx5.87$. Extrema for $r_{\min}/r_0\approx0.20$ and $r_{\max}/r_0\approx8.11$. Inset: Zoom on the local maximum.}
\label{fig:Veff}
\end{figure*}

The effective potential $V_\text{eff}$ can be given in a complicated yet analytical form that we do not report here, but its plot is shown in \cref{fig:Veff}, for both the black-hole and the star-like branch.

\subsection{Black-hole branch}

In the black-hole branch, the effective potential is bounded, vanishes at the horizon and at infinity as $1/r^3$.
The typical behaviour is shown in the left panel of \cref{fig:Veff}.
For $c_1/r_0\leq-2$, it is always negative while for $c_1/r_0>-2$, it develops a principal potential barrier near the horizon and a secondary smaller barrier for larger~$r$ (see the inset of the figure).
As $c_1/r_0$ gets closer to zero, the principal potential barrier becomes higher and narrower while the horizon and the position of the maximum coincide and shrink to zero.

Before integrating numerically \cref{eqpert}, a first hint on the (in)stability of this black-hole spacetime comes from the Simon's criterion~\cite{Simon:1976un}:
a necessary but not sufficient condition for the absence of bound states with $\omega^2<0$ is that
\be\label{simon}
S \equiv \int_{-\infty}^{+\infty} V_\text{eff}\,dr_* > 0\,.
\ee%
It turns out that this quantity is positive for $c_1/r_0>\Lambda_\text{S}$, with $\Lambda_\text{S}\approx-0.4788$,
and therefore we limit our stability analysis within this region.

To show that the black-hole background solution is linearly stable, we need to show that there are no solutions to \cref{eqpert} with $\omega^2<0$ satisfying appropriate boundary conditions.
Because of the behaviour of the potential at the horizon and at infinity, the solution behaves as a purely outgoing (ingoing) free-wave at infinity (horizon), \ie\ $\psi\sim e^{\pm i\omega r_*}$.
For values of $\Lambda_\text{S}<c_1/r_0<0$, we integrate numerically \cref{eqpert} with $\omega^2$ negative but smaller than the depth of the well, and we shoot for the value of $\psi$ and its derivative on the peak of the effective potential in order to have exponentially decreasing solutions on the horizon.
For $c_1/r_0\gtrsim-0.05$ the numerical integration becomes particularly challenging.
If each mode blows up at infinity, independently on the value of $\omega^2$, the perturbation equation does not have bound states.
In our analysis we always find a bound state, then we conclude that the black-hole branch is unstable against linear perturbations.
One remark is, however, in order: this instability is somehow marginal as there is only a finite number of unstable modes.

\subsection{Star-like branch}

In the star-like branch, the asymptotic behaviour of the effective potential is $V_\text{eff}\sim{2r_0/r^3}$ as $r\to\infty$ while near the origin it diverges as $V_\text{eff}\sim{r_0^2/64r^4}$ as $r\to0$. For $0.14\lesssim r/r_0 \lesssim5.87$ it is negative while for $r\gtrsim5.87$ it is positive and has a local maximum for $r/r_0\approx8.11$.
Its plot is shown in the right panel of \cref{fig:Veff}.

Again, to show that the background solution is linearly stable, we need to show that there are no solutions to \cref{eqpert} with $\omega^2<0$ satisfying appropriate boundary conditions.
At spatial infinity we can use purely outgoing, free-wave, boundary conditions, \ie\ $\psi_\infty\sim e^{i\omega r_*}$.
Boundary conditions near the origin are more complicated, due to the behaviour of $V_\text{eff}$ near $r=0$.
More technically, $r=0$ is a non-Fuchsian point and as a consequence, the solution $\psi_0$ near the origin is not a polynomial.
\Cref{eqpert} cannot be solved in terms of simple functions in this limit for any $\omega^2$, nevertheless, for marginally stable solutions ($\omega^2=0$) the solution behaves as $\psi_0(\omega^2=0)\sim e^{-r_0/r}/r$.
For this reason we expect that $\psi_0$ must also be exponentially suppressed for $\omega^2\neq0$.

Because of the very steep barrier at the origin, neither the Simon's criterion nor an S-deformation method~\cite{Kimura:2017uor} are applicable.

In addition, both the barrier at the origin and the lack of more precise boundary conditions near the origin make the numerical integration of \cref{eqpert} very challenging.
For some values of the parameters it is possible to find solutions to \cref{eqpert} for negative and positive values of~$\omega^2$, but such results are highly dependent on the initial parameters. More importantly, we had difficulty in keeping control on the numerical error which (generically) grows of several orders of magnitude at $r\approx r_{\min}$.
For these reasons, we cannot state whether or not the background solution is stable against linear perturbations.
However, although the limit $c_1/r_0\to0$ of the black-hole branch is singular, the instability of the black-hole background solution suggests instability also for the star-like branch.
Yet, the instability time scale could be extremely large (even larger than the Hubble time) and the sine-Gordon solitonic scalar star may still have astrophysical interest.

\section{Discussion}

In this work we have introduced an exact, analytic, static, spherically symmetric, four-dimensional solution of minimally coupled Einstein-scalar gravity sourced by a sine-Gordon scalar soliton. Depending on the value of the parameter $c_1/r_0$, it describes either a black hole or a star-like solution that we called sine-Gordon solitonic scalar star. The scalar potential is not given \emph{a priori} but it is determined by the field equations.

The black hole is characterised by a positive mass and the corresponding scalar potential is bounded, although it always exhibits a negative region.
We have shown that this spacetime is unstable against linear perturbations and we have used it as an effective description to investigate the stability of the star-like solution in the $c_1/r_0\to0$ limit.

The sine-Gordon solitonic scalar star is a horizonless, everywhere regular, asymptotically flat spacetime with positive mass and compactness of $\O(0.1)$.
The scalar potential behaves as a negative constant near the origin and goes to zero as $\phi^5$ at spatial infinity.
Likewise, the energy density of the solution is negative and finite near the origin, becomes positive at a certain radius and vanishes in the asymptotic region.
In that sense, this solution interpolates from the AdS spacetime near the origin and the Schwarzschild spacetime at spatial infinity. 

This peculiar behaviour resembles that of gravastars~\cite{Mazur:2001fv}.
These exotic compact objects have been proposed as alternatives to black holes~\cite{Visser:2003ge} and they are objects whose interior is described by a patch of de Sitter space (characterised by negative pressure) smoothly connected to the Schwarzschild exterior through an intermediate region filled with some (exotic) matter.
In analogy with the gravastar picture, our solution can be regarded as an \emph{anti}-gravastar or the string-inspired AdS bubbles~\cite{Danielsson:2017riq}.
The advantage with respect to these models is that our solution does not require junction conditions with the drawback of a very complicated scalar potential. Notice, however, that our solution is not as compact as a typical gravastar.

The solution-generating method introduced and the result discussed in this work bode well for a possible analytical interpolating solution between de Sitter and Schwarzschild spacetimes, but its search is left for future work.

Unfortunately, we were not able to determine with confidence whether or not the background solution is stable against linear perturbations.
Because of the form of the effective potential, the study of linear perturbations is indeed very complicated both analytically and numerically.
This kind of solutions are often plagued by instabilities~\cite{Cardoso:2007az,Cardoso:2014sna} and probably a full numerical simulation is required.
Similar solitonic solutions sourced by negative energy densities obtained numerically with a Higgs-like scalar potential were shown to be linearly unstable~\cite{Kleihaus:2013tba}.
In addition, results on the black-hole branch may suggest linear instability also in the star-like branch.
However, the number of unstable modes in the black-hole branch is finite and the instability time scale could be sufficiently large to let the sine-Gordon solitonic scalar star still have some astrophysical interest.

Another interesting point that we have not investigated here is the formation mechanism of such a solution.
While the solitonic nature of the scalar profile is comprehensible, the origin of the scalar potential is more mysterious.
Again, a full numerical study of gravitational collapse of scalar matter should be necessary to completely answer this question.

\bigskip\acknowledgments%
We are grateful to Masashi Kimura, Carlos Herdeiro and Eugen Radu for correspondence, to Paolo Pani for discussions and useful comments on a draft of this manuscript, and to Ivica Smoli\'c for a careful reading and comments on a draft of this manuscript.
EF acknowledges financial support from the Angelo Della Riccia Foundation and the Albert Einstein Institute of Potsdam for hospitality during which part of this work was carried out.

\bibliographystyle{apsrev4-1}
\bibliography{refs}

\begin{thebibliography}{57}%
\makeatletter
\providecommand \@ifxundefined [1]{%
 \@ifx{#1\undefined}
}%
\providecommand \@ifnum [1]{%
 \ifnum #1\expandafter \@firstoftwo
 \else \expandafter \@secondoftwo
 \fi
}%
\providecommand \@ifx [1]{%
 \ifx #1\expandafter \@firstoftwo
 \else \expandafter \@secondoftwo
 \fi
}%
\providecommand \natexlab [1]{#1}%
\providecommand \enquote  [1]{``#1''}%
\providecommand \bibnamefont  [1]{#1}%
\providecommand \bibfnamefont [1]{#1}%
\providecommand \citenamefont [1]{#1}%
\providecommand \href@noop [0]{\@secondoftwo}%
\providecommand \href [0]{\begingroup \@sanitize@url \@href}%
\providecommand \@href[1]{\@@startlink{#1}\@@href}%
\providecommand \@@href[1]{\endgroup#1\@@endlink}%
\providecommand \@sanitize@url [0]{\catcode `\\12\catcode `\$12\catcode
  `\&12\catcode `\#12\catcode `\^12\catcode `\_12\catcode `\%12\relax}%
\providecommand \@@startlink[1]{}%
\providecommand \@@endlink[0]{}%
\providecommand \url  [0]{\begingroup\@sanitize@url \@url }%
\providecommand \@url [1]{\endgroup\@href {#1}{\urlprefix }}%
\providecommand \urlprefix  [0]{URL }%
\providecommand \Eprint [0]{\href }%
\providecommand \doibase [0]{http://dx.doi.org/}%
\providecommand \selectlanguage [0]{\@gobble}%
\providecommand \bibinfo  [0]{\@secondoftwo}%
\providecommand \bibfield  [0]{\@secondoftwo}%
\providecommand \translation [1]{[#1]}%
\providecommand \BibitemOpen [0]{}%
\providecommand \bibitemStop [0]{}%
\providecommand \bibitemNoStop [0]{.\EOS\space}%
\providecommand \EOS [0]{\spacefactor3000\relax}%
\providecommand \BibitemShut  [1]{\csname bibitem#1\endcsname}%
\let\auto@bib@innerbib\@empty
\bibitem [{\citenamefont {Ade}\ \emph {et~al.}(2016)\citenamefont {Ade} \emph
  {et~al.}}]{Ade:2015xua}%
  \BibitemOpen
  \bibfield  {author} {\bibinfo {author} {\bibfnamefont {P.~A.~R.}\
  \bibnamefont {Ade}} \emph {et~al.} (\bibinfo {collaboration} {Planck}),\
  }\href {\doibase 10.1051/0004-6361/201525830} {\bibfield  {journal} {\bibinfo
   {journal} {Astron. Astrophys.}\ }\textbf {\bibinfo {volume} {594}},\
  \bibinfo {pages} {A13} (\bibinfo {year} {2016})},\ \Eprint
  {http://arxiv.org/abs/1502.01589} {arXiv:1502.01589} \BibitemShut {NoStop}%
\bibitem [{\citenamefont {Milgrom}(1983)}]{Milgrom:1983ca}%
  \BibitemOpen
  \bibfield  {author} {\bibinfo {author} {\bibfnamefont {M.}~\bibnamefont
  {Milgrom}},\ }\href {\doibase 10.1086/161130} {\bibfield  {journal} {\bibinfo
   {journal} {Astrophys. J.}\ }\textbf {\bibinfo {volume} {270}},\ \bibinfo
  {pages} {365} (\bibinfo {year} {1983})}\BibitemShut {NoStop}%
\bibitem [{\citenamefont {Bertone}(2010)}]{Bertone:2010}%
  \BibitemOpen
  \bibinfo {editor} {\bibfnamefont {G.}~\bibnamefont {Bertone}},\ ed.,\
  \href@noop {} {\emph {\bibinfo {title} {{Particle Dark Matter: Observations,
  Models and Searches}}}}\ (\bibinfo  {publisher} {Cambridge University
  Press},\ \bibinfo {address} {Cambridge, England},\ \bibinfo {year}
  {2010})\BibitemShut {NoStop}%
\bibitem [{\citenamefont {Verlinde}(2017)}]{Verlinde:2016toy}%
  \BibitemOpen
  \bibfield  {author} {\bibinfo {author} {\bibfnamefont {E.~P.}\ \bibnamefont
  {Verlinde}},\ }\href {\doibase 10.21468/SciPostPhys.2.3.016} {\bibfield
  {journal} {\bibinfo  {journal} {SciPost Phys.}\ }\textbf {\bibinfo {volume}
  {2}},\ \bibinfo {pages} {016} (\bibinfo {year} {2017})},\ \Eprint
  {http://arxiv.org/abs/1611.02269} {arXiv:1611.02269} \BibitemShut {NoStop}%
\bibitem [{\citenamefont {Cadoni}\ \emph
  {et~al.}(2018{\natexlab{a}})\citenamefont {Cadoni}, \citenamefont {Casadio},
  \citenamefont {Giusti}, \citenamefont {M{\"u}ck},\ and\ \citenamefont
  {Tuveri}}]{Cadoni:2017evg}%
  \BibitemOpen
  \bibfield  {author} {\bibinfo {author} {\bibfnamefont {M.}~\bibnamefont
  {Cadoni}}, \bibinfo {author} {\bibfnamefont {R.}~\bibnamefont {Casadio}},
  \bibinfo {author} {\bibfnamefont {A.}~\bibnamefont {Giusti}}, \bibinfo
  {author} {\bibfnamefont {W.}~\bibnamefont {M{\"u}ck}}, \ and\ \bibinfo
  {author} {\bibfnamefont {M.}~\bibnamefont {Tuveri}},\ }\href {\doibase
  10.1016/j.physletb.2017.11.058} {\bibfield  {journal} {\bibinfo  {journal}
  {Phys. Lett.}\ }\textbf {\bibinfo {volume} {B776}},\ \bibinfo {pages} {242}
  (\bibinfo {year} {2018}{\natexlab{a}})},\ \Eprint
  {http://arxiv.org/abs/1707.09945} {arXiv:1707.09945} \BibitemShut {NoStop}%
\bibitem [{\citenamefont {Cadoni}\ \emph
  {et~al.}(2018{\natexlab{b}})\citenamefont {Cadoni}, \citenamefont {Casadio},
  \citenamefont {Giusti},\ and\ \citenamefont {Tuveri}}]{Cadoni:2018dnd}%
  \BibitemOpen
  \bibfield  {author} {\bibinfo {author} {\bibfnamefont {M.}~\bibnamefont
  {Cadoni}}, \bibinfo {author} {\bibfnamefont {R.}~\bibnamefont {Casadio}},
  \bibinfo {author} {\bibfnamefont {A.}~\bibnamefont {Giusti}}, \ and\ \bibinfo
  {author} {\bibfnamefont {M.}~\bibnamefont {Tuveri}},\ }\href {\doibase
  10.1103/PhysRevD.97.044047} {\bibfield  {journal} {\bibinfo  {journal} {Phys.
  Rev.}\ }\textbf {\bibinfo {volume} {D97}},\ \bibinfo {pages} {044047}
  (\bibinfo {year} {2018}{\natexlab{b}})},\ \Eprint
  {http://arxiv.org/abs/1801.10374} {arXiv:1801.10374} \BibitemShut {NoStop}%
\bibitem [{\citenamefont {Marsh}\ and\ \citenamefont
  {Pop}(2015)}]{Marsh:2015wka}%
  \BibitemOpen
  \bibfield  {author} {\bibinfo {author} {\bibfnamefont {D.~J.~E.}\
  \bibnamefont {Marsh}}\ and\ \bibinfo {author} {\bibfnamefont {A.-R.}\
  \bibnamefont {Pop}},\ }\href {\doibase 10.1093/mnras/stv1050} {\bibfield
  {journal} {\bibinfo  {journal} {Mon. Notices Royal Astron. Soc.}\ }\textbf
  {\bibinfo {volume} {451}},\ \bibinfo {pages} {2479} (\bibinfo {year}
  {2015})},\ \Eprint {http://arxiv.org/abs/1502.03456} {arXiv:1502.03456}
  \BibitemShut {NoStop}%
\bibitem [{\citenamefont {Schunck}\ and\ \citenamefont
  {Mielke}(2003)}]{Schunck:2003kk}%
  \BibitemOpen
  \bibfield  {author} {\bibinfo {author} {\bibfnamefont {F.~E.}\ \bibnamefont
  {Schunck}}\ and\ \bibinfo {author} {\bibfnamefont {E.~W.}\ \bibnamefont
  {Mielke}},\ }\href {\doibase 10.1088/0264-9381/20/20/201} {\bibfield
  {journal} {\bibinfo  {journal} {Class. Quantum Grav.}\ }\textbf {\bibinfo
  {volume} {20}},\ \bibinfo {pages} {R301} (\bibinfo {year} {2003})},\ \Eprint
  {http://arxiv.org/abs/0801.0307} {arXiv:0801.0307} \BibitemShut {NoStop}%
\bibitem [{\citenamefont {Liebling}\ and\ \citenamefont
  {Palenzuela}(2017)}]{Liebling:2012fv}%
  \BibitemOpen
  \bibfield  {author} {\bibinfo {author} {\bibfnamefont {S.~L.}\ \bibnamefont
  {Liebling}}\ and\ \bibinfo {author} {\bibfnamefont {C.}~\bibnamefont
  {Palenzuela}},\ }\href {\doibase 10.1007/s41114-017-0007-y} {\bibfield
  {journal} {\bibinfo  {journal} {Living Rev. Relativ.}\ }\textbf {\bibinfo
  {volume} {20}},\ \bibinfo {pages} {5} (\bibinfo {year} {2017})},\ \Eprint
  {http://arxiv.org/abs/1202.5809} {arXiv:1202.5809} \BibitemShut {NoStop}%
\bibitem [{\citenamefont {Mielke}\ and\ \citenamefont
  {Schunck}(2000)}]{Mielke:2000mh}%
  \BibitemOpen
  \bibfield  {author} {\bibinfo {author} {\bibfnamefont {E.~W.}\ \bibnamefont
  {Mielke}}\ and\ \bibinfo {author} {\bibfnamefont {F.~E.}\ \bibnamefont
  {Schunck}},\ }\href {\doibase 10.1016/S0550-3213(99)00492-7} {\bibfield
  {journal} {\bibinfo  {journal} {Nucl. Phys.}\ }\textbf {\bibinfo {volume}
  {B564}},\ \bibinfo {pages} {185} (\bibinfo {year} {2000})},\ \Eprint
  {http://arxiv.org/abs/gr-qc/0001061} {arXiv:gr-qc/0001061} \BibitemShut
  {NoStop}%
\bibitem [{\citenamefont {Guzm\'an}\ and\ \citenamefont
  {Rueda-Becerril}(2009)}]{Guzman:2009zz}%
  \BibitemOpen
  \bibfield  {author} {\bibinfo {author} {\bibfnamefont {F.~S.}\ \bibnamefont
  {Guzm\'an}}\ and\ \bibinfo {author} {\bibfnamefont {J.~M.}\ \bibnamefont
  {Rueda-Becerril}},\ }\href {\doibase 10.1103/PhysRevD.80.084023} {\bibfield
  {journal} {\bibinfo  {journal} {Phys. Rev.}\ }\textbf {\bibinfo {volume}
  {D80}},\ \bibinfo {pages} {084023} (\bibinfo {year} {2009})},\ \Eprint
  {http://arxiv.org/abs/1009.1250} {arXiv:1009.1250} \BibitemShut {NoStop}%
\bibitem [{\citenamefont {Macedo}\ \emph
  {et~al.}(2013{\natexlab{a}})\citenamefont {Macedo}, \citenamefont {Pani},
  \citenamefont {Cardoso},\ and\ \citenamefont {Crispino}}]{Macedo:2013jja}%
  \BibitemOpen
  \bibfield  {author} {\bibinfo {author} {\bibfnamefont {C.~F.~B.}\
  \bibnamefont {Macedo}}, \bibinfo {author} {\bibfnamefont {P.}~\bibnamefont
  {Pani}}, \bibinfo {author} {\bibfnamefont {V.}~\bibnamefont {Cardoso}}, \
  and\ \bibinfo {author} {\bibfnamefont {L.~C.~B.}\ \bibnamefont {Crispino}},\
  }\href {\doibase 10.1103/PhysRevD.88.064046} {\bibfield  {journal} {\bibinfo
  {journal} {Phys. Rev.}\ }\textbf {\bibinfo {volume} {D88}},\ \bibinfo {pages}
  {064046} (\bibinfo {year} {2013}{\natexlab{a}})},\ \Eprint
  {http://arxiv.org/abs/1307.4812} {arXiv:1307.4812} \BibitemShut {NoStop}%
\bibitem [{\citenamefont {Barausse}\ \emph {et~al.}(2014)\citenamefont
  {Barausse}, \citenamefont {Cardoso},\ and\ \citenamefont
  {Pani}}]{Barausse:2014tra}%
  \BibitemOpen
  \bibfield  {author} {\bibinfo {author} {\bibfnamefont {E.}~\bibnamefont
  {Barausse}}, \bibinfo {author} {\bibfnamefont {V.}~\bibnamefont {Cardoso}}, \
  and\ \bibinfo {author} {\bibfnamefont {P.}~\bibnamefont {Pani}},\ }\href
  {\doibase 10.1103/PhysRevD.89.104059} {\bibfield  {journal} {\bibinfo
  {journal} {Phys. Rev.}\ }\textbf {\bibinfo {volume} {D89}},\ \bibinfo {pages}
  {104059} (\bibinfo {year} {2014})},\ \Eprint {http://arxiv.org/abs/1404.7149}
  {arXiv:1404.7149} \BibitemShut {NoStop}%
\bibitem [{\citenamefont {Vincent}\ \emph {et~al.}(2016)\citenamefont
  {Vincent}, \citenamefont {Meliani}, \citenamefont {Grandcl\'ement},
  \citenamefont {Gourgoulhon},\ and\ \citenamefont {Straub}}]{Vincent:2015xta}%
  \BibitemOpen
  \bibfield  {author} {\bibinfo {author} {\bibfnamefont {F.~H.}\ \bibnamefont
  {Vincent}}, \bibinfo {author} {\bibfnamefont {Z.}~\bibnamefont {Meliani}},
  \bibinfo {author} {\bibfnamefont {P.}~\bibnamefont {Grandcl\'ement}},
  \bibinfo {author} {\bibfnamefont {E.}~\bibnamefont {Gourgoulhon}}, \ and\
  \bibinfo {author} {\bibfnamefont {O.}~\bibnamefont {Straub}},\ }\href
  {\doibase 10.1088/0264-9381/33/10/105015} {\bibfield  {journal} {\bibinfo
  {journal} {Class. Quantum Grav.}\ }\textbf {\bibinfo {volume} {33}},\
  \bibinfo {pages} {105015} (\bibinfo {year} {2016})},\ \Eprint
  {http://arxiv.org/abs/1510.04170} {arXiv:1510.04170} \BibitemShut {NoStop}%
\bibitem [{\citenamefont {Macedo}\ \emph
  {et~al.}(2013{\natexlab{b}})\citenamefont {Macedo}, \citenamefont {Pani},
  \citenamefont {Cardoso},\ and\ \citenamefont {Crispino}}]{Macedo:2013qea}%
  \BibitemOpen
  \bibfield  {author} {\bibinfo {author} {\bibfnamefont {C.~F.~B.}\
  \bibnamefont {Macedo}}, \bibinfo {author} {\bibfnamefont {P.}~\bibnamefont
  {Pani}}, \bibinfo {author} {\bibfnamefont {V.}~\bibnamefont {Cardoso}}, \
  and\ \bibinfo {author} {\bibfnamefont {L.~C.~B.}\ \bibnamefont {Crispino}},\
  }\href {\doibase 10.1088/0004-637X/774/1/48} {\bibfield  {journal} {\bibinfo
  {journal} {Astrophys. J.}\ }\textbf {\bibinfo {volume} {774}},\ \bibinfo
  {pages} {48} (\bibinfo {year} {2013}{\natexlab{b}})},\ \Eprint
  {http://arxiv.org/abs/1302.2646} {arXiv:1302.2646} \BibitemShut {NoStop}%
\bibitem [{\citenamefont {Cardoso}\ \emph
  {et~al.}(2016{\natexlab{a}})\citenamefont {Cardoso}, \citenamefont
  {Franzin},\ and\ \citenamefont {Pani}}]{Cardoso:2016rao}%
  \BibitemOpen
  \bibfield  {author} {\bibinfo {author} {\bibfnamefont {V.}~\bibnamefont
  {Cardoso}}, \bibinfo {author} {\bibfnamefont {E.}~\bibnamefont {Franzin}}, \
  and\ \bibinfo {author} {\bibfnamefont {P.}~\bibnamefont {Pani}},\ }\href
  {\doibase 10.1103/PhysRevLett.116.171101} {\bibfield  {journal} {\bibinfo
  {journal} {Phys. Rev. Lett.}\ }\textbf {\bibinfo {volume} {116}},\ \bibinfo
  {pages} {171101} (\bibinfo {year} {2016}{\natexlab{a}})},\ \bibinfo {note}
  {{Erratum: Phys. Rev. Lett. {\bf117}, 089902(E) (2016)}},\ \Eprint
  {http://arxiv.org/abs/1602.07309} {arXiv:1602.07309} \BibitemShut {NoStop}%
\bibitem [{\citenamefont {Cardoso}\ \emph
  {et~al.}(2016{\natexlab{b}})\citenamefont {Cardoso}, \citenamefont {Hopper},
  \citenamefont {Macedo}, \citenamefont {Palenzuela},\ and\ \citenamefont
  {Pani}}]{Cardoso:2016oxy}%
  \BibitemOpen
  \bibfield  {author} {\bibinfo {author} {\bibfnamefont {V.}~\bibnamefont
  {Cardoso}}, \bibinfo {author} {\bibfnamefont {S.}~\bibnamefont {Hopper}},
  \bibinfo {author} {\bibfnamefont {C.~F.~B.}\ \bibnamefont {Macedo}}, \bibinfo
  {author} {\bibfnamefont {C.}~\bibnamefont {Palenzuela}}, \ and\ \bibinfo
  {author} {\bibfnamefont {P.}~\bibnamefont {Pani}},\ }\href {\doibase
  10.1103/PhysRevD.94.084031} {\bibfield  {journal} {\bibinfo  {journal} {Phys.
  Rev.}\ }\textbf {\bibinfo {volume} {D94}},\ \bibinfo {pages} {084031}
  (\bibinfo {year} {2016}{\natexlab{b}})},\ \Eprint
  {http://arxiv.org/abs/1608.08637} {arXiv:1608.08637} \BibitemShut {NoStop}%
\bibitem [{\citenamefont {Cardoso}\ \emph {et~al.}(2017)\citenamefont
  {Cardoso}, \citenamefont {Franzin}, \citenamefont {Maselli}, \citenamefont
  {Pani},\ and\ \citenamefont {Raposo}}]{Cardoso:2017cfl}%
  \BibitemOpen
  \bibfield  {author} {\bibinfo {author} {\bibfnamefont {V.}~\bibnamefont
  {Cardoso}}, \bibinfo {author} {\bibfnamefont {E.}~\bibnamefont {Franzin}},
  \bibinfo {author} {\bibfnamefont {A.}~\bibnamefont {Maselli}}, \bibinfo
  {author} {\bibfnamefont {P.}~\bibnamefont {Pani}}, \ and\ \bibinfo {author}
  {\bibfnamefont {G.}~\bibnamefont {Raposo}},\ }\href {\doibase
  10.1103/PhysRevD.95.084014} {\bibfield  {journal} {\bibinfo  {journal} {Phys.
  Rev.}\ }\textbf {\bibinfo {volume} {D95}},\ \bibinfo {pages} {084014}
  (\bibinfo {year} {2017})},\ \Eprint {http://arxiv.org/abs/1701.01116}
  {arXiv:1701.01116} \BibitemShut {NoStop}%
\bibitem [{\citenamefont {Sennett}\ \emph {et~al.}(2017)\citenamefont
  {Sennett}, \citenamefont {Hinderer}, \citenamefont {Steinhoff}, \citenamefont
  {Buonanno},\ and\ \citenamefont {Ossokine}}]{Sennett:2017etc}%
  \BibitemOpen
  \bibfield  {author} {\bibinfo {author} {\bibfnamefont {N.}~\bibnamefont
  {Sennett}}, \bibinfo {author} {\bibfnamefont {T.}~\bibnamefont {Hinderer}},
  \bibinfo {author} {\bibfnamefont {J.}~\bibnamefont {Steinhoff}}, \bibinfo
  {author} {\bibfnamefont {A.}~\bibnamefont {Buonanno}}, \ and\ \bibinfo
  {author} {\bibfnamefont {S.}~\bibnamefont {Ossokine}},\ }\href {\doibase
  10.1103/PhysRevD.96.024002} {\bibfield  {journal} {\bibinfo  {journal} {Phys.
  Rev.}\ }\textbf {\bibinfo {volume} {D96}},\ \bibinfo {pages} {024002}
  (\bibinfo {year} {2017})},\ \Eprint {http://arxiv.org/abs/1704.08651}
  {arXiv:1704.08651} \BibitemShut {NoStop}%
\bibitem [{\citenamefont {Seidel}\ and\ \citenamefont
  {Suen}(1991)}]{Seidel:1991zh}%
  \BibitemOpen
  \bibfield  {author} {\bibinfo {author} {\bibfnamefont {E.}~\bibnamefont
  {Seidel}}\ and\ \bibinfo {author} {\bibfnamefont {W.~M.}\ \bibnamefont
  {Suen}},\ }\href {\doibase 10.1103/PhysRevLett.66.1659} {\bibfield  {journal}
  {\bibinfo  {journal} {Phys. Rev. Lett.}\ }\textbf {\bibinfo {volume} {66}},\
  \bibinfo {pages} {1659} (\bibinfo {year} {1991})}\BibitemShut {NoStop}%
\bibitem [{\citenamefont {Bekenstein}(1972)}]{Bekenstein:1971hc}%
  \BibitemOpen
  \bibfield  {author} {\bibinfo {author} {\bibfnamefont {J.~D.}\ \bibnamefont
  {Bekenstein}},\ }\href {\doibase 10.1103/PhysRevD.5.1239} {\bibfield
  {journal} {\bibinfo  {journal} {Phys. Rev.}\ }\textbf {\bibinfo {volume}
  {D5}},\ \bibinfo {pages} {1239} (\bibinfo {year} {1972})}\BibitemShut
  {NoStop}%
\bibitem [{\citenamefont {Bekenstein}(1995)}]{Bekenstein:1995un}%
  \BibitemOpen
  \bibfield  {author} {\bibinfo {author} {\bibfnamefont {J.~D.}\ \bibnamefont
  {Bekenstein}},\ }\href {\doibase 10.1103/PhysRevD.51.R6608} {\bibfield
  {journal} {\bibinfo  {journal} {Phys. Rev.}\ }\textbf {\bibinfo {volume}
  {D51}},\ \bibinfo {pages} {R6608} (\bibinfo {year} {1995})}\BibitemShut
  {NoStop}%
\bibitem [{\citenamefont {Sudarsky}(1995)}]{Sudarsky:1995zg}%
  \BibitemOpen
  \bibfield  {author} {\bibinfo {author} {\bibfnamefont {D.}~\bibnamefont
  {Sudarsky}},\ }\href {\doibase 10.1088/0264-9381/12/2/023} {\bibfield
  {journal} {\bibinfo  {journal} {Class. Quantum Grav.}\ }\textbf {\bibinfo
  {volume} {12}},\ \bibinfo {pages} {579} (\bibinfo {year} {1995})}\BibitemShut
  {NoStop}%
\bibitem [{\citenamefont {Torii}\ \emph {et~al.}(2001)\citenamefont {Torii},
  \citenamefont {Maeda},\ and\ \citenamefont {Narita}}]{Torii:2001pg}%
  \BibitemOpen
  \bibfield  {author} {\bibinfo {author} {\bibfnamefont {T.}~\bibnamefont
  {Torii}}, \bibinfo {author} {\bibfnamefont {K.}~\bibnamefont {Maeda}}, \ and\
  \bibinfo {author} {\bibfnamefont {M.}~\bibnamefont {Narita}},\ }\href
  {\doibase 10.1103/PhysRevD.64.044007} {\bibfield  {journal} {\bibinfo
  {journal} {Phys. Rev.}\ }\textbf {\bibinfo {volume} {D64}},\ \bibinfo {pages}
  {044007} (\bibinfo {year} {2001})}\BibitemShut {NoStop}%
\bibitem [{\citenamefont {Hertog}(2006)}]{Hertog:2006rr}%
  \BibitemOpen
  \bibfield  {author} {\bibinfo {author} {\bibfnamefont {T.}~\bibnamefont
  {Hertog}},\ }\href {\doibase 10.1103/PhysRevD.74.084008} {\bibfield
  {journal} {\bibinfo  {journal} {Phys. Rev.}\ }\textbf {\bibinfo {volume}
  {D74}},\ \bibinfo {pages} {084008} (\bibinfo {year} {2006})},\ \Eprint
  {http://arxiv.org/abs/gr-qc/0608075} {arXiv:gr-qc/0608075} \BibitemShut
  {NoStop}%
\bibitem [{\citenamefont {Herdeiro}\ and\ \citenamefont
  {Radu}(2014{\natexlab{a}})}]{Herdeiro:2014goa}%
  \BibitemOpen
  \bibfield  {author} {\bibinfo {author} {\bibfnamefont {C.~A.~R.}\
  \bibnamefont {Herdeiro}}\ and\ \bibinfo {author} {\bibfnamefont
  {E.}~\bibnamefont {Radu}},\ }\href {\doibase 10.1103/PhysRevLett.112.221101}
  {\bibfield  {journal} {\bibinfo  {journal} {Phys. Rev. Lett.}\ }\textbf
  {\bibinfo {volume} {112}},\ \bibinfo {pages} {221101} (\bibinfo {year}
  {2014}{\natexlab{a}})},\ \Eprint {http://arxiv.org/abs/1403.2757}
  {arXiv:1403.2757} \BibitemShut {NoStop}%
\bibitem [{\citenamefont {Herdeiro}\ and\ \citenamefont
  {Radu}(2014{\natexlab{b}})}]{Herdeiro:2015waa}%
  \BibitemOpen
  \bibfield  {author} {\bibinfo {author} {\bibfnamefont {C.~A.~R.}\
  \bibnamefont {Herdeiro}}\ and\ \bibinfo {author} {\bibfnamefont
  {E.}~\bibnamefont {Radu}},\ }in\ \href {\doibase 10.1142/S0218271815420146}
  {\emph {\bibinfo {booktitle} {{Proceedings of the 7th Black Holes
  Workshop}}}},\ \bibinfo {series and number} {\bibinfo {number} {Int.\ J.\
  Mod.\ Phys.\ {\bf D24}}},\ \bibinfo {editor} {edited by\ \bibinfo {editor}
  {\bibfnamefont {C.~A.~R.}\ \bibnamefont {Herdeiro}}, \bibinfo {editor}
  {\bibfnamefont {V.}~\bibnamefont {Cardoso}}, \bibinfo {editor} {\bibfnamefont
  {J.~P.~S.}\ \bibnamefont {Lemos}}, \ and\ \bibinfo {editor} {\bibfnamefont
  {F.~C.}\ \bibnamefont {Mena}}}\ (\bibinfo {address} {Aveiro, Portugal},\
  \bibinfo {year} {Dec.\ 18--19, 2014})\ p.\ \bibinfo {pages} {1542014},\
  \Eprint {http://arxiv.org/abs/1504.08209} {arXiv:1504.08209} \BibitemShut
  {NoStop}%
\bibitem [{\citenamefont {Nucamendi}\ and\ \citenamefont
  {Salgado}(2003)}]{Nucamendi:1995ex}%
  \BibitemOpen
  \bibfield  {author} {\bibinfo {author} {\bibfnamefont {U.}~\bibnamefont
  {Nucamendi}}\ and\ \bibinfo {author} {\bibfnamefont {M.}~\bibnamefont
  {Salgado}},\ }\href {\doibase 10.1103/PhysRevD.68.044026} {\bibfield
  {journal} {\bibinfo  {journal} {Phys. Rev.}\ }\textbf {\bibinfo {volume}
  {D68}},\ \bibinfo {pages} {044026} (\bibinfo {year} {2003})},\ \Eprint
  {http://arxiv.org/abs/gr-qc/0301062} {arXiv:gr-qc/0301062} \BibitemShut
  {NoStop}%
\bibitem [{\citenamefont {Bechmann}\ and\ \citenamefont
  {Lechtenfeld}(1995)}]{Bechmann:1995sa}%
  \BibitemOpen
  \bibfield  {author} {\bibinfo {author} {\bibfnamefont {O.}~\bibnamefont
  {Bechmann}}\ and\ \bibinfo {author} {\bibfnamefont {O.}~\bibnamefont
  {Lechtenfeld}},\ }\href {\doibase 10.1088/0264-9381/12/6/013} {\bibfield
  {journal} {\bibinfo  {journal} {Class. Quantum Grav.}\ }\textbf {\bibinfo
  {volume} {12}},\ \bibinfo {pages} {1473} (\bibinfo {year} {1995})},\ \Eprint
  {http://arxiv.org/abs/gr-qc/9502011} {arXiv:gr-qc/9502011} \BibitemShut
  {NoStop}%
\bibitem [{\citenamefont {Rubinstein}(1970)}]{Rubinstein:1970vb}%
  \BibitemOpen
  \bibfield  {author} {\bibinfo {author} {\bibfnamefont {J.}~\bibnamefont
  {Rubinstein}},\ }\href {\doibase 10.1063/1.1665057} {\bibfield  {journal}
  {\bibinfo  {journal} {J. Math. Phys.}\ }\textbf {\bibinfo {volume} {11}},\
  \bibinfo {pages} {258} (\bibinfo {year} {1970})}\BibitemShut {NoStop}%
\bibitem [{\citenamefont {Gaeta}\ \emph {et~al.}(1994)\citenamefont {Gaeta},
  \citenamefont {Reiss}, \citenamefont {Peyrard},\ and\ \citenamefont
  {Dauxois}}]{Gaeta}%
  \BibitemOpen
  \bibfield  {author} {\bibinfo {author} {\bibfnamefont {G.}~\bibnamefont
  {Gaeta}}, \bibinfo {author} {\bibfnamefont {C.}~\bibnamefont {Reiss}},
  \bibinfo {author} {\bibfnamefont {M.}~\bibnamefont {Peyrard}}, \ and\
  \bibinfo {author} {\bibfnamefont {T.}~\bibnamefont {Dauxois}},\ }\href
  {\doibase 10.1007/BF02724511} {\bibfield  {journal} {\bibinfo  {journal}
  {Riv. Nuovo Cim.}\ }\textbf {\bibinfo {volume} {17}},\ \bibinfo {pages} {1}
  (\bibinfo {year} {1994})}\BibitemShut {NoStop}%
\bibitem [{\citenamefont {Barone}\ and\ \citenamefont
  {Patern\`o}(1982)}]{Barone}%
  \BibitemOpen
  \bibfield  {author} {\bibinfo {author} {\bibfnamefont {A.}~\bibnamefont
  {Barone}}\ and\ \bibinfo {author} {\bibfnamefont {G.}~\bibnamefont
  {Patern\`o}},\ }\href@noop {} {\emph {\bibinfo {title} {{Physics and
  Applications of the Josephson Effect}}}}\ (\bibinfo  {publisher} {Wiley},\
  \bibinfo {address} {New York, USA},\ \bibinfo {year} {1982})\BibitemShut
  {NoStop}%
\bibitem [{\citenamefont {Cuevas-Maraver}\ \emph {et~al.}(2014)\citenamefont
  {Cuevas-Maraver}, \citenamefont {Kevrekidis},\ and\ \citenamefont
  {Williams}}]{Cuevas-Maraver}%
  \BibitemOpen
  \bibinfo {editor} {\bibfnamefont {J.}~\bibnamefont {Cuevas-Maraver}},
  \bibinfo {editor} {\bibfnamefont {P.}~\bibnamefont {Kevrekidis}}, \ and\
  \bibinfo {editor} {\bibfnamefont {F.}~\bibnamefont {Williams}},\ eds.,\
  \href@noop {} {\emph {\bibinfo {title} {{The sine-Gordon Model and its
  Applications}}}}\ (\bibinfo  {publisher} {Springer},\ \bibinfo {address}
  {Heidelberg, Germany},\ \bibinfo {year} {2014})\BibitemShut {NoStop}%
\bibitem [{\citenamefont {Gegenberg}\ and\ \citenamefont
  {Kunstatter}(1997)}]{Gegenberg:1997ns}%
  \BibitemOpen
  \bibfield  {author} {\bibinfo {author} {\bibfnamefont {J.}~\bibnamefont
  {Gegenberg}}\ and\ \bibinfo {author} {\bibfnamefont {G.}~\bibnamefont
  {Kunstatter}},\ }\href {\doibase 10.1016/S0370-2693(97)01118-0} {\bibfield
  {journal} {\bibinfo  {journal} {Phys. Lett.}\ }\textbf {\bibinfo {volume}
  {B413}},\ \bibinfo {pages} {274} (\bibinfo {year} {1997})},\ \Eprint
  {http://arxiv.org/abs/hep-th/9707181} {arXiv:hep-th/9707181} \BibitemShut
  {NoStop}%
\bibitem [{\citenamefont {Cadoni}(1998)}]{Cadoni:1998ej}%
  \BibitemOpen
  \bibfield  {author} {\bibinfo {author} {\bibfnamefont {M.}~\bibnamefont
  {Cadoni}},\ }\href {\doibase 10.1103/PhysRevD.58.104001} {\bibfield
  {journal} {\bibinfo  {journal} {Phys. Rev.}\ }\textbf {\bibinfo {volume}
  {D58}},\ \bibinfo {pages} {104001} (\bibinfo {year} {1998})},\ \Eprint
  {http://arxiv.org/abs/hep-th/9803257} {arXiv:hep-th/9803257} \BibitemShut
  {NoStop}%
\bibitem [{\citenamefont {Su}\ and\ \citenamefont {Yan}(2012)}]{Su:2012}%
  \BibitemOpen
  \bibfield  {author} {\bibinfo {author} {\bibfnamefont {W.-J.}\ \bibnamefont
  {Su}}\ and\ \bibinfo {author} {\bibfnamefont {J.}~\bibnamefont {Yan}},\
  }\href {\doibase 10.1139/p2012-106} {\bibfield  {journal} {\bibinfo
  {journal} {Can. J. Phys.}\ }\textbf {\bibinfo {volume} {90}},\ \bibinfo
  {pages} {1279} (\bibinfo {year} {2012})}\BibitemShut {NoStop}%
\bibitem [{\citenamefont {Cadoni}\ \emph {et~al.}(2011)\citenamefont {Cadoni},
  \citenamefont {Serra},\ and\ \citenamefont {Mignemi}}]{Cadoni:2011nq}%
  \BibitemOpen
  \bibfield  {author} {\bibinfo {author} {\bibfnamefont {M.}~\bibnamefont
  {Cadoni}}, \bibinfo {author} {\bibfnamefont {M.}~\bibnamefont {Serra}}, \
  and\ \bibinfo {author} {\bibfnamefont {S.}~\bibnamefont {Mignemi}},\ }\href
  {\doibase 10.1103/PhysRevD.84.084046} {\bibfield  {journal} {\bibinfo
  {journal} {Phys. Rev.}\ }\textbf {\bibinfo {volume} {D84}},\ \bibinfo {pages}
  {084046} (\bibinfo {year} {2011})},\ \Eprint {http://arxiv.org/abs/1107.5979}
  {arXiv:1107.5979} \BibitemShut {NoStop}%
\bibitem [{\citenamefont {Cadoni}\ \emph {et~al.}(2012)\citenamefont {Cadoni},
  \citenamefont {Serra},\ and\ \citenamefont {Mignemi}}]{Cadoni:2011yj}%
  \BibitemOpen
  \bibfield  {author} {\bibinfo {author} {\bibfnamefont {M.}~\bibnamefont
  {Cadoni}}, \bibinfo {author} {\bibfnamefont {M.}~\bibnamefont {Serra}}, \
  and\ \bibinfo {author} {\bibfnamefont {S.}~\bibnamefont {Mignemi}},\ }\href
  {\doibase 10.1103/PhysRevD.85.086001} {\bibfield  {journal} {\bibinfo
  {journal} {Phys. Rev.}\ }\textbf {\bibinfo {volume} {D85}},\ \bibinfo {pages}
  {086001} (\bibinfo {year} {2012})},\ \Eprint {http://arxiv.org/abs/1111.6581}
  {arXiv:1111.6581} \BibitemShut {NoStop}%
\bibitem [{\citenamefont {Cadoni}\ and\ \citenamefont
  {Mignemi}(2012)}]{Cadoni:2012uf}%
  \BibitemOpen
  \bibfield  {author} {\bibinfo {author} {\bibfnamefont {M.}~\bibnamefont
  {Cadoni}}\ and\ \bibinfo {author} {\bibfnamefont {S.}~\bibnamefont
  {Mignemi}},\ }\href {\doibase 10.1007/JHEP06(2012)056} {\bibfield  {journal}
  {\bibinfo  {journal} {J. High Energy Phys.}\ }\textbf {\bibinfo {volume}
  {06}},\ \bibinfo {pages} {056} (\bibinfo {year} {2012})},\ \Eprint
  {http://arxiv.org/abs/1205.0412} {arXiv:1205.0412} \BibitemShut {NoStop}%
\bibitem [{\citenamefont {Cadoni}\ and\ \citenamefont
  {Serra}(2012)}]{Cadoni:2012ea}%
  \BibitemOpen
  \bibfield  {author} {\bibinfo {author} {\bibfnamefont {M.}~\bibnamefont
  {Cadoni}}\ and\ \bibinfo {author} {\bibfnamefont {M.}~\bibnamefont {Serra}},\
  }\href {\doibase 10.1007/JHEP11(2012)136} {\bibfield  {journal} {\bibinfo
  {journal} {J. High Energy Phys.}\ }\textbf {\bibinfo {volume} {11}},\
  \bibinfo {pages} {136} (\bibinfo {year} {2012})},\ \Eprint
  {http://arxiv.org/abs/1209.4484} {arXiv:1209.4484} \BibitemShut {NoStop}%
\bibitem [{\citenamefont {Cadoni}\ and\ \citenamefont
  {Franzin}(2015)}]{Cadoni:2015gfa}%
  \BibitemOpen
  \bibfield  {author} {\bibinfo {author} {\bibfnamefont {M.}~\bibnamefont
  {Cadoni}}\ and\ \bibinfo {author} {\bibfnamefont {E.}~\bibnamefont
  {Franzin}},\ }\href {\doibase 10.1103/PhysRevD.91.104011} {\bibfield
  {journal} {\bibinfo  {journal} {Phys. Rev.}\ }\textbf {\bibinfo {volume}
  {D91}},\ \bibinfo {pages} {104011} (\bibinfo {year} {2015})},\ \Eprint
  {http://arxiv.org/abs/1503.04734} {arXiv:1503.04734} \BibitemShut {NoStop}%
\bibitem [{\citenamefont {Cadoni}\ \emph {et~al.}(2016)\citenamefont {Cadoni},
  \citenamefont {Franzin},\ and\ \citenamefont {Serra}}]{Cadoni:2015gce}%
  \BibitemOpen
  \bibfield  {author} {\bibinfo {author} {\bibfnamefont {M.}~\bibnamefont
  {Cadoni}}, \bibinfo {author} {\bibfnamefont {E.}~\bibnamefont {Franzin}}, \
  and\ \bibinfo {author} {\bibfnamefont {M.}~\bibnamefont {Serra}},\ }\href
  {\doibase 10.1007/JHEP01(2016)125} {\bibfield  {journal} {\bibinfo  {journal}
  {J. High Energy Phys.}\ }\textbf {\bibinfo {volume} {01}},\ \bibinfo {pages}
  {125} (\bibinfo {year} {2016})},\ \Eprint {http://arxiv.org/abs/1511.03986}
  {arXiv:1511.03986} \BibitemShut {NoStop}%
\bibitem [{\citenamefont {Azreg-A{\"i}nou}(2010)}]{AzregAinou:2009dj}%
  \BibitemOpen
  \bibfield  {author} {\bibinfo {author} {\bibfnamefont {M.}~\bibnamefont
  {Azreg-A{\"i}nou}},\ }\href {\doibase 10.1007/s10714-009-0915-6} {\bibfield
  {journal} {\bibinfo  {journal} {Gen. Rel. Gravit.}\ }\textbf {\bibinfo
  {volume} {42}},\ \bibinfo {pages} {1427} (\bibinfo {year} {2010})},\ \Eprint
  {http://arxiv.org/abs/0912.1722} {arXiv:0912.1722} \BibitemShut {NoStop}%
\bibitem [{\citenamefont {Solovyev}\ and\ \citenamefont
  {Tsirulev}(2012)}]{Solovyev:2012zz}%
  \BibitemOpen
  \bibfield  {author} {\bibinfo {author} {\bibfnamefont {D.~A.}\ \bibnamefont
  {Solovyev}}\ and\ \bibinfo {author} {\bibfnamefont {A.~N.}\ \bibnamefont
  {Tsirulev}},\ }\href {\doibase 10.1088/0264-9381/29/5/055013} {\bibfield
  {journal} {\bibinfo  {journal} {Class. Quantum Grav.}\ }\textbf {\bibinfo
  {volume} {29}},\ \bibinfo {pages} {055013} (\bibinfo {year}
  {2012})}\BibitemShut {NoStop}%
\bibitem [{\citenamefont {Smoli\'c}(2015)}]{Smolic:2015txa}%
  \BibitemOpen
  \bibfield  {author} {\bibinfo {author} {\bibfnamefont {I.}~\bibnamefont
  {Smoli\'c}},\ }\href {\doibase 10.1088/0264-9381/32/14/145010} {\bibfield
  {journal} {\bibinfo  {journal} {Class. Quantum Grav.}\ }\textbf {\bibinfo
  {volume} {32}},\ \bibinfo {pages} {145010} (\bibinfo {year} {2015})},\
  \Eprint {http://arxiv.org/abs/1501.04967} {arXiv:1501.04967} \BibitemShut
  {NoStop}%
\bibitem [{\citenamefont {Smoli\'c}(2017)}]{Smolic:2016dmh}%
  \BibitemOpen
  \bibfield  {author} {\bibinfo {author} {\bibfnamefont {I.}~\bibnamefont
  {Smoli\'c}},\ }\href {\doibase 10.1103/PhysRevD.95.024016} {\bibfield
  {journal} {\bibinfo  {journal} {Phys. Rev.}\ }\textbf {\bibinfo {volume}
  {D95}},\ \bibinfo {pages} {024016} (\bibinfo {year} {2017})},\ \Eprint
  {http://arxiv.org/abs/1609.04013} {arXiv:1609.04013} \BibitemShut {NoStop}%
\bibitem [{\citenamefont {Amaro-Seoane}\ \emph {et~al.}(2010)\citenamefont
  {Amaro-Seoane}, \citenamefont {Barranco}, \citenamefont {Bernal},\ and\
  \citenamefont {Rezzol\-la}}]{AmaroSeoane:2010qx}%
  \BibitemOpen
  \bibfield  {author} {\bibinfo {author} {\bibfnamefont {P.}~\bibnamefont
  {Amaro-Seoane}}, \bibinfo {author} {\bibfnamefont {J.}~\bibnamefont
  {Barranco}}, \bibinfo {author} {\bibfnamefont {A.}~\bibnamefont {Bernal}}, \
  and\ \bibinfo {author} {\bibfnamefont {L.}~\bibnamefont {Rezzol\-la}},\
  }\href {\doibase 10.1088/1475-7516/2010/11/002} {\bibfield  {journal}
  {\bibinfo  {journal} {J. Cosmol. Astropart. Phys.}\ }\textbf {\bibinfo
  {volume} {11}},\ \bibinfo {pages} {002} (\bibinfo {year} {2010})},\ \Eprint
  {http://arxiv.org/abs/1009.0019} {arXiv:1009.0019} \BibitemShut {NoStop}%
\bibitem [{\citenamefont {Madsen}(1985)}]{Madsen:1985}%
  \BibitemOpen
  \bibfield  {author} {\bibinfo {author} {\bibfnamefont {M.~S.}\ \bibnamefont
  {Madsen}},\ }\href {\doibase 10.1007/BF00650285} {\bibfield  {journal}
  {\bibinfo  {journal} {Astrophys. Space Sci.}\ }\textbf {\bibinfo {volume}
  {113}},\ \bibinfo {pages} {205} (\bibinfo {year} {1985})}\BibitemShut
  {NoStop}%
\bibitem [{\citenamefont {Bronnikov}\ \emph {et~al.}(2011)\citenamefont
  {Bronnikov}, \citenamefont {Fabris},\ and\ \citenamefont
  {Zhidenko}}]{Bronnikov:2011if}%
  \BibitemOpen
  \bibfield  {author} {\bibinfo {author} {\bibfnamefont {K.~A.}\ \bibnamefont
  {Bronnikov}}, \bibinfo {author} {\bibfnamefont {J.~C.}\ \bibnamefont
  {Fabris}}, \ and\ \bibinfo {author} {\bibfnamefont {A.}~\bibnamefont
  {Zhidenko}},\ }\href {\doibase 10.1140/epjc/s10052-011-1791-2} {\bibfield
  {journal} {\bibinfo  {journal} {Eur. Phys. J.}\ }\textbf {\bibinfo {volume}
  {C71}},\ \bibinfo {pages} {1791} (\bibinfo {year} {2011})},\ \Eprint
  {http://arxiv.org/abs/1109.6576} {arXiv:1109.6576} \BibitemShut {NoStop}%
\bibitem [{\citenamefont {Simon}(1976)}]{Simon:1976un}%
  \BibitemOpen
  \bibfield  {author} {\bibinfo {author} {\bibfnamefont {B.}~\bibnamefont
  {Simon}},\ }\href@noop {} {\bibfield  {journal} {\bibinfo  {journal} {Ann.
  Phys. (N. Y.)}\ }\textbf {\bibinfo {volume} {97}},\ \bibinfo {pages} {279}
  (\bibinfo {year} {1976})}\BibitemShut {NoStop}%
\bibitem [{\citenamefont {Kimura}(2017)}]{Kimura:2017uor}%
  \BibitemOpen
  \bibfield  {author} {\bibinfo {author} {\bibfnamefont {M.}~\bibnamefont
  {Kimura}},\ }\href {\doibase 10.1088/1361-6382/aa903f} {\bibfield  {journal}
  {\bibinfo  {journal} {Class. Quantum Grav.}\ }\textbf {\bibinfo {volume}
  {34}},\ \bibinfo {pages} {235007} (\bibinfo {year} {2017})},\ \Eprint
  {http://arxiv.org/abs/1706.01447} {arXiv:1706.01447} \BibitemShut {NoStop}%
\bibitem [{\citenamefont {Mazur}\ and\ \citenamefont
  {Mottola}(2001)}]{Mazur:2001fv}%
  \BibitemOpen
  \bibfield  {author} {\bibinfo {author} {\bibfnamefont {P.~O.}\ \bibnamefont
  {Mazur}}\ and\ \bibinfo {author} {\bibfnamefont {E.}~\bibnamefont
  {Mottola}},\ }\href@noop {} {\  (\bibinfo {year} {2001})},\ \Eprint
  {http://arxiv.org/abs/gr-qc/0109035} {arXiv:gr-qc/0109035} \BibitemShut
  {NoStop}%
\bibitem [{\citenamefont {Visser}\ and\ \citenamefont
  {Wiltshire}(2004)}]{Visser:2003ge}%
  \BibitemOpen
  \bibfield  {author} {\bibinfo {author} {\bibfnamefont {M.}~\bibnamefont
  {Visser}}\ and\ \bibinfo {author} {\bibfnamefont {D.~L.}\ \bibnamefont
  {Wiltshire}},\ }\href {\doibase 10.1088/0264-9381/21/4/027} {\bibfield
  {journal} {\bibinfo  {journal} {Class. Quantum Grav.}\ }\textbf {\bibinfo
  {volume} {21}},\ \bibinfo {pages} {1135} (\bibinfo {year} {2004})},\ \Eprint
  {http://arxiv.org/abs/gr-qc/0310107} {arXiv:gr-qc/0310107} \BibitemShut
  {NoStop}%
\bibitem [{\citenamefont {Danielsson}\ \emph {et~al.}(2017)\citenamefont
  {Danielsson}, \citenamefont {Dibitetto},\ and\ \citenamefont
  {Giri}}]{Danielsson:2017riq}%
  \BibitemOpen
  \bibfield  {author} {\bibinfo {author} {\bibfnamefont {U.~H.}\ \bibnamefont
  {Danielsson}}, \bibinfo {author} {\bibfnamefont {G.}~\bibnamefont
  {Dibitetto}}, \ and\ \bibinfo {author} {\bibfnamefont {S.}~\bibnamefont
  {Giri}},\ }\href {\doibase 10.1007/JHEP10(2017)171} {\bibfield  {journal}
  {\bibinfo  {journal} {J. High Energy Phys.}\ }\textbf {\bibinfo {volume}
  {10}},\ \bibinfo {pages} {171} (\bibinfo {year} {2017})},\ \Eprint
  {http://arxiv.org/abs/1705.10172} {arXiv:1705.10172} \BibitemShut {NoStop}%
\bibitem [{\citenamefont {Cardoso}\ \emph {et~al.}(2008)\citenamefont
  {Cardoso}, \citenamefont {Pani}, \citenamefont {Cadoni},\ and\ \citenamefont
  {Cavagli\`a}}]{Cardoso:2007az}%
  \BibitemOpen
  \bibfield  {author} {\bibinfo {author} {\bibfnamefont {V.}~\bibnamefont
  {Cardoso}}, \bibinfo {author} {\bibfnamefont {P.}~\bibnamefont {Pani}},
  \bibinfo {author} {\bibfnamefont {M.}~\bibnamefont {Cadoni}}, \ and\ \bibinfo
  {author} {\bibfnamefont {M.}~\bibnamefont {Cavagli\`a}},\ }\href {\doibase
  10.1103/PhysRevD.77.124044} {\bibfield  {journal} {\bibinfo  {journal} {Phys.
  Rev.}\ }\textbf {\bibinfo {volume} {D77}},\ \bibinfo {pages} {124044}
  (\bibinfo {year} {2008})},\ \Eprint {http://arxiv.org/abs/0709.0532}
  {arXiv:0709.0532} \BibitemShut {NoStop}%
\bibitem [{\citenamefont {Cardoso}\ \emph {et~al.}(2014)\citenamefont
  {Cardoso}, \citenamefont {Crispino}, \citenamefont {Macedo}, \citenamefont
  {Okawa},\ and\ \citenamefont {Pani}}]{Cardoso:2014sna}%
  \BibitemOpen
  \bibfield  {author} {\bibinfo {author} {\bibfnamefont {V.}~\bibnamefont
  {Cardoso}}, \bibinfo {author} {\bibfnamefont {L.~C.~B.}\ \bibnamefont
  {Crispino}}, \bibinfo {author} {\bibfnamefont {C.~F.~B.}\ \bibnamefont
  {Macedo}}, \bibinfo {author} {\bibfnamefont {H.}~\bibnamefont {Okawa}}, \
  and\ \bibinfo {author} {\bibfnamefont {P.}~\bibnamefont {Pani}},\ }\href
  {\doibase 10.1103/PhysRevD.90.044069} {\bibfield  {journal} {\bibinfo
  {journal} {Phys. Rev.}\ }\textbf {\bibinfo {volume} {D90}},\ \bibinfo {pages}
  {044069} (\bibinfo {year} {2014})},\ \Eprint {http://arxiv.org/abs/1406.5510}
  {arXiv:1406.5510} \BibitemShut {NoStop}%
\bibitem [{\citenamefont {Kleihaus}\ \emph {et~al.}(2013)\citenamefont
  {Kleihaus}, \citenamefont {Kunz}, \citenamefont {Radu},\ and\ \citenamefont
  {Subagyo}}]{Kleihaus:2013tba}%
  \BibitemOpen
  \bibfield  {author} {\bibinfo {author} {\bibfnamefont {B.}~\bibnamefont
  {Kleihaus}}, \bibinfo {author} {\bibfnamefont {J.}~\bibnamefont {Kunz}},
  \bibinfo {author} {\bibfnamefont {E.}~\bibnamefont {Radu}}, \ and\ \bibinfo
  {author} {\bibfnamefont {B.}~\bibnamefont {Subagyo}},\ }\href {\doibase
  10.1016/j.physletb.2013.07.051} {\bibfield  {journal} {\bibinfo  {journal}
  {Phys. Lett.}\ }\textbf {\bibinfo {volume} {B725}},\ \bibinfo {pages} {489}
  (\bibinfo {year} {2013})},\ \Eprint {http://arxiv.org/abs/1306.4616}
  {arXiv:1306.4616} \BibitemShut {NoStop}%
\end{thebibliography}%

\end{document}